\documentclass[useAMS,usenatbib]{mnras}
\usepackage[T1]{fontenc}
\usepackage{ae,aecompl}
\usepackage{pdflscape}
\usepackage{graphicx}
\usepackage{longtable}
\usepackage{lscape}
\usepackage{amssymb}
\usepackage{endnotes} 
\usepackage{footnote}
\usepackage{amsmath}
\usepackage{txfonts}
\usepackage{natbib}
\usepackage{url}
\usepackage{breakurl}
\usepackage{helvet}
\usepackage{rotating}
\usepackage{tikz}
\usepackage{float}
\usepackage{lipsum}
\usepackage{isotope}
\usepackage{supertabular}
\usepackage{setspace}
\usepackage{hyperref}
\usepackage{caption}
\usepackage{subcaption}

\title[SN~2017hpa]{SN~2017hpa: A carbon-rich type Ia supernova}
\author[Dutta et al.]{
Anirban Dutta$^{1}$\thanks{E-mail: anirban.dutta@iiap.res.in},
Avinash Singh$^{2}$,
G. C. Anupama$^{1}$,
D. K. Sahu$^{1}$ and
Brajesh Kumar$^{2}$
\\
$^{1}$Indian Institute of Astrophysics, II Block, Koramangala, Bangalore 560 034, India\\
$^{2}$Aryabhatta Research Institute of Observational Sciences, Manora Peak, Nainital, 263 001 India\\
}
\date{Accepted ------------, Received ------------; in original form ------------}
\pagerange{\pageref{firstpage}--\pageref{lastpage}} \pubyear{}

\begin{document}
\label{firstpage}
\pagerange{\pageref{firstpage}--\pageref{lastpage}}
\maketitle
% Abstract of the paper
\begin{abstract}
We present the optical (\textit{UBVRI}) and ultraviolet (\textit{Swift}-UVOT) photometry, and optical spectroscopy of Type Ia supernova SN~2017hpa. We study broadband UV+optical light curves and low resolution spectroscopy spanning from $-13.8$ to $+108$~d from the maximum light in $B$-band. The photometric analysis indicates that SN~2017hpa is a normal type Ia with \(\Delta m_{B}(15) = 0.98\pm0.16\) mag and \(M_{B}=-19.45\pm0.15\) mag at a distance modulus of \(\mu = 34.08\pm0.09\) mag. The $(uvw1-uvv)$ colour evolution shows that SN~2017hpa falls in the NUV-blue group. The $(B-V)$ colour at maximum is bluer in comparison to normal type Ia supernovae. Spectroscopic analysis shows that the Si\,{\sc ii} 6355 absorption feature evolves rapidly with a velocity gradient, \(\dot{v}=128\pm 7\) km s$^{-1}$\,d$^{-1}$. The pre-maximum phase spectra show prominent C\,{\sc ii} 6580 \text{\AA} absorption feature. The C\,{\sc ii} 6580 \text{\AA} line velocity measured from the observed spectra is lower than the velocity of Si\,{\sc ii} 6355 \text{\AA}, which could be due to a line of sight effect. The synthetic spectral fits to the pre-maximum spectra using \textit{syn++} indicate the presence of a high velocity component in the Si\,{\sc ii} absorption, in addition to a photospheric component. Fitting the observed spectrum with the spectral synthesis code {\small \textit{TARDIS}}, the mass of unburned C in the ejecta is estimated to be $\sim 0.019$~$M_{\odot}$. The peak bolometric luminosity is \(L^{\rm{bol}}_{\rm{peak}} = 1.43\times10^{43}\) erg s$^{-1}$. The radiation diffusion model fit to the bolometric light curve indicates \(0.61\pm0.02\) \(M_\odot\) of $^{56}$Ni is synthesized in the explosion. 

\end{abstract}

% Select between one and six entries from the list of approved keywords.
% Don't make up new ones.

%%%%%%%%%%%%%%%%%%%%%%%%%%%%%%%%%%%%%%%%%%%%%%%%%%

%%%%%%%%%%%%%%%%% BODY OF PAPER %%%%%%%%%%%%%%%%%%

\section{Introduction}

Type Ia supernovae (SNe Ia) are widely regarded as explosions due to thermonuclear runaway in a white dwarf (WD), most likely carbon-oxygen with a mass close to the Chandrasekhar limit \citep{1960ApJ...132..565H}. However, specific progenitor systems and mechanisms driving the explosions have not been clearly identified. Two of the most favoured models for the progenitors are the single degenerate (SD) systems, \citep[]{ 1982ApJ...253..798N, 1982ApJ...257..780N, 1982ApJ...253..785A, 1984ApJS...54..335I} and the double-degenerate (DD) systems \citep[]{1984ApJS...54..335I,1984ApJ...277..355W}. In the former case, the WD accretes matter from a non-degenerate companion (giant, main-sequence or helium star) in a close binary system, while in the latter case, the explosion is produced by the merger of two white dwarfs (\citealt{2010ApJ...719.1067K, 2013ApJ...770L...8P}). The best-studied models so far involve a Chandrasekhar mass WD explosion \citep{2009MNRAS.399L.156R}, but recent studies of sub-Chandrasekhar mass models (\citealt{2012MNRAS.420.3003S, 2018ApJ...852L..33G}) have been successful in reproducing the peak brightness and light curve widths of the fast-declining SNe Ia.

A reasonably tight correlation between the luminosity at peak and the light curve decline rate has been established for a majority of normal SNe Ia (\citealt{1993ApJ...413L.105P}, \citealt{1999AJ....118.1766P}), with brighter objects having a slower decline rate. This correlation allows the estimation of precise distances to these objects, and hence the measurement for the expansion rate of the Universe \citep{1996ApJ...473...88R, 1998AJ....116.1009R, 1998ApJ...504..935R, 1999AJ....118.1766P, 1999ApJ...517..565P}. 

There exists a spectroscopic classification \citep{1995ApJ...455L.147N} defined by the ratio of the equivalent widths of the absorption features of Si\,{\sc ii} at $\lambda$5972 and $\lambda$6355. This ratio, R(Si II), is related with the absolute magnitude of the supernova and hence with the decline rate. It acts as a distance independent luminosity indicator. This is interpreted as due to temperature difference and also the variation in the amount of $\rm ^{56}Ni$ produced in the supernova explosion. Based on the ratio of the pseudo-equivalent widths (pEW) of Si\,{\sc ii} $\lambda$5972 and $\lambda$6355 features in the spectra near maximum light, SNe Ia can be grouped into core normal (CN), shallow silicon (SS), broadline (BL) and cool (CL) classes \citep{2006PASP..118..560B}. This indicates inhomogeneities in the compositional structure and temperature difference in SNe Ia.

\cite{2005ApJ...623.1011B} provided a further sub-classification based on the evolution of the expansion velocity, in which the objects fall under three groups: FAINT, High-Velocity Gradient (HVG), and Low-Velocity Gradient (LVG). There are some hints that the HVG objects could result from delayed detonation, while the LVG objects could be due to deflagration (\citealt{2000ApJ...530..966L, 2005ApJ...623.1011B}). This spectral diversity can also be explained by considering an asymmetric explosion viewed from different line of sight \citep{2010Natur.466...82M}.

In the pre-maximum phase, spectra of SNe Ia consist of overlapping P-Cygni profiles of intermediate-mass elements (IME's) like Si, S, Mg, Ca, and Iron group elements (IGE's) like Fe, Co, Ni. In some cases, carbon and oxygen features are present in the very early spectra, which could result from incomplete burning (\citealt{2011ApJ...732...30P, 2012ApJ...745...74F, 2014MNRAS.444.3258M}). The expansion velocity inferred using these features could provide an essential hint to the velocity structure of the outer expanding ejecta. It is not certain as to whether the unburned material is present only in the outer layers or is mixed within the ejecta, and if so, the extent of mixing.
The objects with detection of unburned material are of great importance as they can constrain the amount of material that is still unburned and hence the explosion channel. The pure deflagration model (W7) \citep{1984ApJ...286..644N}, and the pulsating delayed detonation models (\citealt{1995ApJ...444..831H, 2014MNRAS.441..532D}) leave unburned carbon in the ejecta, while the delayed detonation models burn most of the carbon (\citealt{1991A&A...245L..25K, 2009Natur.460..869K}).

Detailed photometric and spectroscopic study of type Ia supernova SN~2017hpa that showed unburned carbon in its early spectrum are presented in this work. The observed properties of SN~2017hpa are compared with those of normal type Ia supernovae, and its explosion parameters are estimated. Supernova SN~2017hpa\footnote{\url{https://wis-tns.weizmann.ac.il/object/2017hpa}} was discovered by \cite{2017TNSTR1164....1G} on 2017 October 25, 08:18:16 UT (JD=2458051.84) in the galaxy UGC 3122\footnote{\url{http://leda.univ-lyon1.fr/}}. A spectrum obtained on 2017 October 25 23:55:02 UT with the Asiago 1.82 m Copernico Telescope equipped with AFOSC was found to be consistent with the very early spectrum of type Ia SN, in particular, SN~1990N $\sim14$ days before the maximum light \citep{2017ATel10896....1F}. Using a spectroscopic redshift of z = 0.0156 for the host galaxy, an expansion velocity of $\sim$16,000 km s$^{-1}$ was deduced from the absorption minimum of Si\,{\sc ii} $\lambda$6355. The important parameters of SN~2017hpa and its host galaxy UGC~3122 are presented in Table~\ref{tab:SN_params}. 

Details of the observations and data reduction methods are presented in section \ref{Observations}. The photometric and spectroscopic analysis are discussed in sections \ref{Light Curve Analysis} and \ref{Spectroscopy}, respectively. Finally, we summarise our results in section \ref{Summary}. 

\section{Observations and Data Reductions} \label{Observations} 
\subsection{Optical photometry with 2-m HCT}
    Bessell $UBVRI$ photometric observations of SN~2017hpa were carried out using the Hanle Faint Object Spectrograph Camera (HFOSC), mounted on the 2-m Himalayan Chandra Telescope (HCT) of the Indian Astronomical Observatory (IAO)\footnote{\url{https://www.iiap.res.in/?q=telescope_iao}}, Hanle, India. The HFOSC is equipped with a $2K \times 4K$ SITe CCD chip, and the central $2K \times 2K$ pixels are used for imaging, covering a field of view (FOV) of $10 \times 10$ arcmin at an image scale of 0.296 arcsec pixel$^{-1}$. The readout noise and gain of the camera are 4.87$e^{-}$ and 1.22$e^{-}$/ADU, respectively. The pre-processing tasks of bias subtraction and flat field correction were performed using the bias and sky flat frames obtained on each night. Cosmic-ray events were removed. All tasks were performed using the standard packages available with {\small \it IRAF}. Whenever multiple frames were observed in the same band, they were combined before performing photometry.
    
\begin{figure}
\centering
\includegraphics[width=\columnwidth]{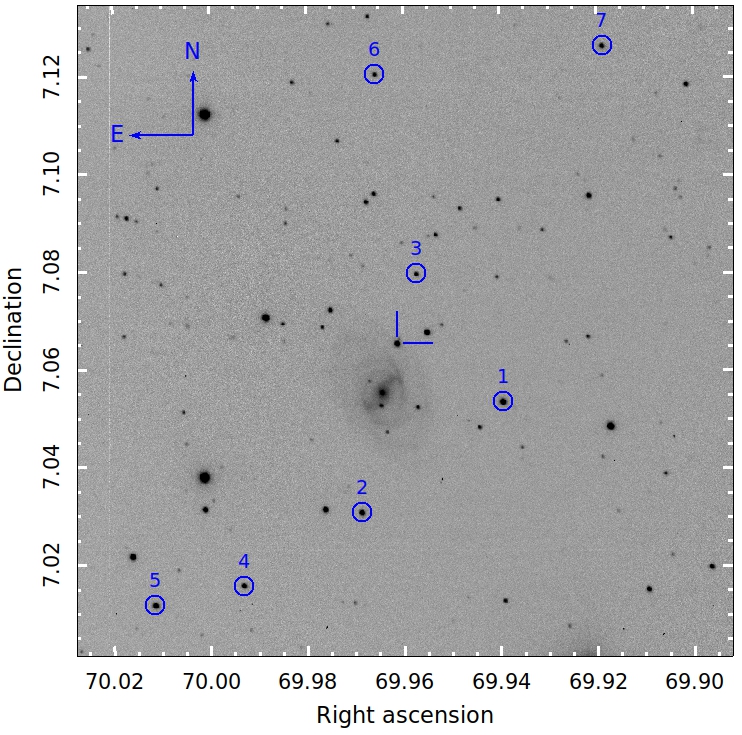}
\caption{SN~2017hpa in the host galaxy UGC 3122. This is a $\sim$\,7\,$\times$\,7 arcmin$^2$ image in $V$-band (50 sec exposure) taken with HCT on 2017 October 31. The stars circled in blue (Ids 1--7) are the secondary standard stars used for calibration. The supernova is marked with crosshairs.}
\label{Fig1}
\end{figure}

    For calibrating a sequence of secondary stars in the SN field, photometric standard stars from the list of Landolt \citep{1992AJ....104..340L} were observed along with the SN field on photometric nights. Standard field PG0231+051 was observed on 2017 November 22 and 2018 January 5, while the field PG0918+029 was observed on 2017 December 29. The magnitudes of the standard stars in the Landolt field and the stars in the SN field were estimated using aperture photometry within the DAOPHOT \citep{1987PASP...99..191S} task in {\small \it IRAF}. The average atmospheric extinction coefficients for the site, in each of the passbands $U$, $B$, $V$, $R$, $I$ were adopted from \cite{2008BASI...36..111S} and secondary stars in the field were calibrated as mentioned in \citet{2018MNRAS.480.2475S}. Fig.~\ref{Fig1} shows the field of SN~2017hpa with the secondary standards identified. The $UBVRI$ magnitudes of the local standard stars are listed in Table~\ref{tab:standardlog}.
    
    The apparent magnitudes of the secondary standard stars were obtained by performing point-spread function (PSF) photometry. The zero points for each night were estimated using the average colour terms for the telescope-detector system. Though the SN lies in the outskirts of its host galaxy, there is a possibility of SN flux contamination from the host and a nearby field star (see Fig.~\ref{Fig1}). Therefore, to remove any possible contamination and estimate the supernova brightness accurately, template subtraction methodology was adopted. The template images of SN~2017hpa were obtained on 2019 September 23 under good seeing conditions with the 2-m HCT. The templates were astrometrically aligned with the science images (containing the SN), background subtracted, PSF matched, and scaled. The scaled template was then subtracted from the science images, leaving only the supernova in the resultant images. Aperture photometry of the supernova was then performed, and the supernova magnitudes were calibrated using the nightly zero points. Final (template subtracted) SN magnitudes are listed in Table~\ref{tab:photlog} and used for further analysis in the paper.
        
\subsection{UV-Optical Photometry with $Swift-UVOT$}
SN~2017hpa was also observed with the {\it Ultra Violet Optical Telescope (UVOT)} onboard the Neil Gehrels {\it Swift} Observatory. The data were obtained from the {\it Swift} archive (\url{https://www.swift.ac.uk/swift_portal/}). The {\it UVOT} observations were made with  broadband filters $uvw2$ (1928 \text{\AA}), $uvm2$ (2246 \text{\AA}), $uvw1$ (2600 \text{\AA}), $u$ (3465 \text{\AA}), $b$ (4392 \text{\AA}) and $v$ (5468 \text{\AA}) starting from 2017 October 26 (JD 2458053.2) and continued till 2017 December 07 (JD 2458095.4). The data reduction was performed using  various packages available with HEASOFT (High Energy Astrophysics Software, version 6.27) and with the latest calibration database for the {\it UVOT} instrument, following the methods as described in \cite{2008MNRAS.383..627P} and \cite{2009AJ....137.4517B}. The SN was not detected in the $uvw2$ and $uvm2$ band with considerable signal-to-noise. The supernova magnitude was extracted using {\small UVOTSOURCE} task with an aperture size of 5$^{''}$ for the source and a similar aperture size to extract the background counts. The final {\it UVOT} magnitudes are obtained in the the Vega system and tabulated in Table~\ref{tab:uvlog}.
    
\subsection{Spectroscopy}    
    The spectroscopic monitoring of SN~2017hpa with the  HCT started on 2017 October 30 (JD 2458057.4) and continued till 2018 February 25 (JD 2458175.1). Low resolution spectra were obtained using grisms, Gr7 (3500-7800 \text{\AA}) and Gr8 (5200-9100 \text{\AA}) available with the HFOSC. The log of spectroscopic observations is provided in Table~\ref{tab:speclog}. The two-dimensional images were pre-processed in a standard way. One-dimensional spectra were optimally extracted \citep{1986PASP...98..609H} from the processed two-dimensional images using the TWODSPEC package in {\small IRAF}. Arc lamp spectra FeAr and FeNe were used for wavelength calibration. Night sky emission lines were used to check the wavelength calibration, and whenever necessary, small shifts were applied. The wavelength calibrated spectra were corrected for the instrumental response using spectra of spectrophotometric standards and brought to a relative flux scale. On the nights when spectrophotometric standards were not observed, response curves from adjacent nights were used. The flux calibrated spectra in the two grisms were combined, scaled to a weighted mean, to give the final spectrum on a relative flux scale. Finally, the combined spectra were scaled with the broadband $UBVRI$ magnitudes to bring it to an absolute flux scale. The spectra have been corrected for redshift (z = 0.0156). Telluric features have been removed from the spectra by using the {\small \it{telluric}} package in {\small IRAF}.

\begin{table}
\caption{Parameters of SN2017hpa and its host galaxy.}
\label{tab:SN_params}
\setlength{\tabcolsep}{2.5pt}
\begin{tabular}{l l c}
\hline \hline 
\noalign{\smallskip}
Parameters 			    & Value		 				                        & Ref.\\

\noalign{\smallskip} \hline \noalign{\smallskip} 
\multicolumn{3}{l}{\textit{SN2017hpa}:} \\ \\

RA (J2000) 			    & $\alpha\,=\,04^{\rm h} 39^{\rm m} 50\fs73$ 	    & 2 \\
DEC (J2000)             & $\delta\,=\,+07\degr 03\arcmin 55\farcs22$ 	    & 2 \\
Galactocentric Location	& 11\farcs2 W, 35\farcs6 N 			                & 2 \\ \\

Discovery Date 		    & $t_{\rm d}$\,=\,2017 October 25 08:18 (UTC) 		& 2 \\
			            & (JD 2458051.84)            			            &   \\
                			            
Date of $B$-band Maxima   & $t_{\rm 0}$\,=\,2017 November 08 17:45 (UTC) 		& 1 \\
			            & (JD 2458066.29$\,\pm\,$0.11)    			        & \\ \\
$\Delta$m$_{15}$($B$)       & $0.98\pm0.16$ mag                             & 1 \\	
Galaxy reddening        & $E(B-V)$\,=\,0.1518$\,\pm\,$0.0069 mag            & 3 \\
Host reddening          & $E(B-V)$\,=\,0.08$\,\pm\,$0.06 mag                & 1 \\
$(B-V)_{0}$              & $-0.26\pm0.03$                                   & 1 \\
Peak Magnitude ($B$-band)	& $M_{B}$\,=\,$-$19.45$\,\pm\,$0.15 mag             & 1 \\
Distance modulus        & $\mu$\,=\,34.08$\,\pm\,$0.09 mag 			        & 1 \\ 
Peak Luminosity         & $L^{\rm{bol}}_{\rm{peak}}$ $=$ 1.43\(\times 10^{43}\) erg s$^{-1}$          & 1 \\
$^{56}$Ni mass          & $M_{Ni}$ $=$ 0.61$\pm$0.02 \(M_\odot\)            & \\
Ejected mass            & $M_{ej}$ $=$ 1.10$\pm$0.22 \(M_\odot\)            & \\
$\dot{v}$               & 127.9 $\pm$ 6.1 km s$^{-1}$\,d$^{-1}$                & 1 \\
R(Si II)$_{max}$        & 0.13\,$\pm$\,0.02                                    & 1 \\
v$_{max}$               & 9643\,$\pm$\,110 km s$^{-1}$                           & 1 \\
v$_{10}$                & 8320\,$\pm$\,120 km s$^{-1}$                           & 1 \\
Kinetic energy          & $E_{K}$ $=$ (0.80\,$\pm$\,0.23)\(\times 10^{51}\) erg $s^{-1}$        & 1 \\

\noalign{\smallskip} \hline \noalign{\smallskip}
\multicolumn{3}{l}{\textit{UGC 3122}:} \\ \\

Alternate name 		    & MCG+01-12-013, PGC15760,                              \\
                        &  MCG1-12-013, CGCG419-21                          & 4 \\
Type                    & SBc 						                        & 4 \\
RA (J2000) 			    & $\alpha\,=\,04^{\rm h} 39^{\rm m} 51\fs50$ 	    & 4 \\
DEC (J2000)             & $\delta\,=\,+07\degr 03\arcmin 19\farcs0$ 	    & 4 \\
Red-shift               & z\,=\,0.015647$\,\pm\,$0.000027                   & 4 \\
\noalign{\smallskip} \hline
\end{tabular}
\newline 
(1) This paper;
(2) \cite{2017TNSTR1164....1G}
(3) \cite{2011ApJ...737..103S}
(4) \cite{2006AJ....131.1163S}
\end{table}

\section{Light Curve}
\label{Light Curve Analysis}
\subsection{Light Curve Analysis}

\begin{figure}
\centering
\includegraphics[width=\columnwidth]{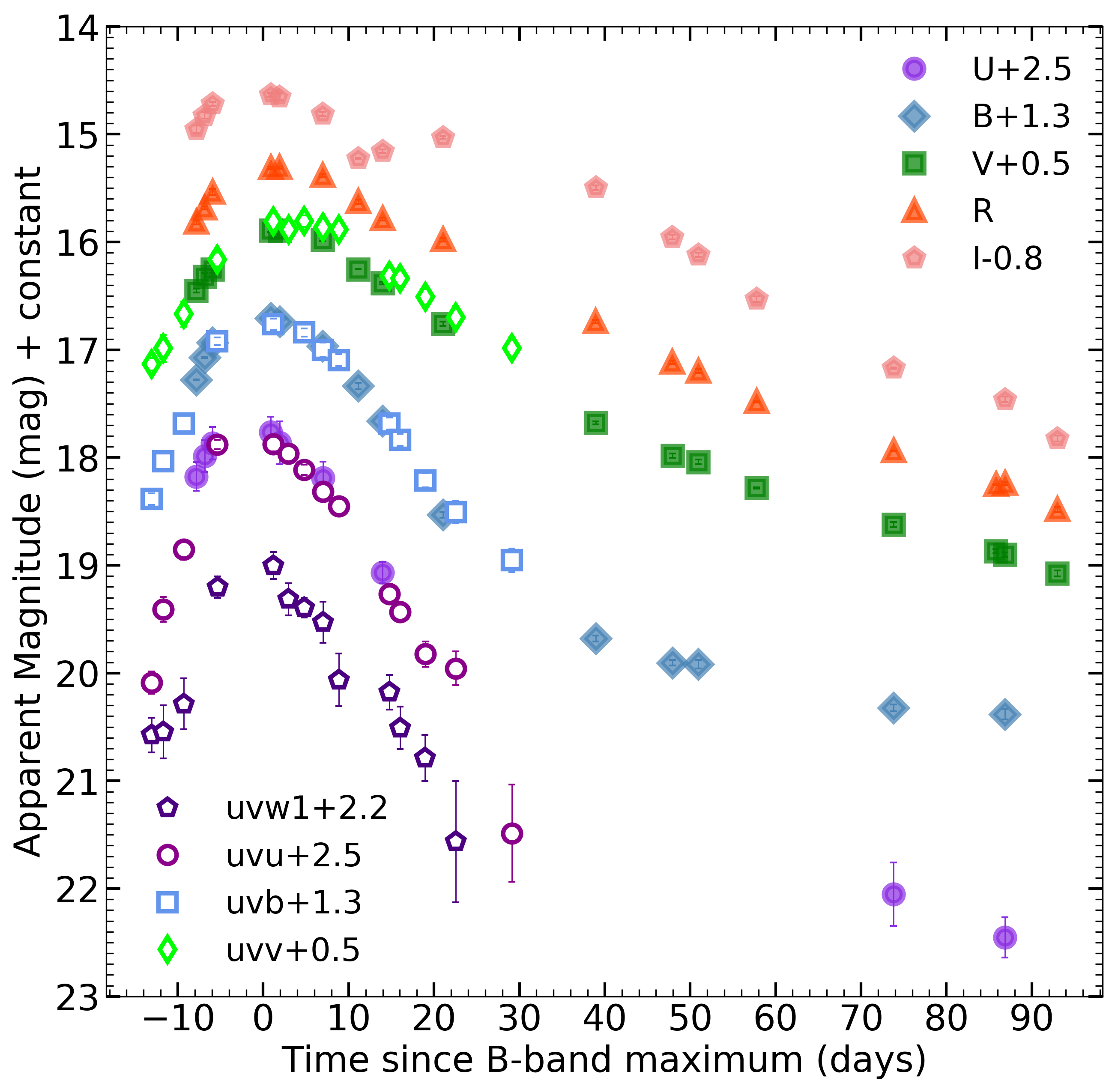}
\caption{$UBVRI$ and $Swift-UVOT$ light curves of SN~2017hpa. The phase is measured with respect to $B$ band maximum. The light curves in individual bands have been shifted for representation purpose as indicated in the legend.}
\label{Fig2}
\end{figure}

The light curves of SN~2017hpa in $U, B, V, R, I$, and $uvw1$, $uvu$, $uvv$, and $uvb$ bands are plotted in Fig.~\ref{Fig2}. The light curve in $U, B, V, R$ and $I$ bands  were fit with MLCS2k2 \citep{2007ApJ...659..122J} (details of fit provided in Sec.~\ref{Extinction}). The light curve parameters, e.g., the decline rate parameter ($\Delta m_{15}(B)$), time of maximum, and the magnitude at maximum have been calculated from the fitted model, and the associated errors are the observed error propagated with the model error. The $B$ band peak magnitude is $15.48\pm0.11$ mag, and  $\Delta m_{15}(B)$=$0.98\pm0.15$ mag. The $I$ band light curve shows a distinct secondary peak characteristic of normal SNe Ia. An inflation is also seen in the $R$ band light curve around the same epoch. The maximum in the $U$ and $I$ band light curve occurs at 2.03 d and 3.04 d before $B$ band maximum, respectively, and that of $V$ and $R$ band occurs around $\sim$ 1.02 day after $B$ band maximum. The early occurrence of the $I$-band maximum is consistent with other well-observed SNe Ia \citep{1999AJ....118.1766P, 2005A&A...429..667A}. The secondary maximum in the $I$-band light curve occurred at +30.46 d after the $B$ band maximum, with a magnitude $0.44$ mag fainter than the peak. The double-peaked nature is directly related to the ionisation evolution of Iron Group Elements (IGE's) in the supernova ejecta, \citep{2006ApJ...649..939K}. As the ejecta expands, it cools down, and at a temperature $T\sim 7000$~K, the near-infrared emission of Fe/Co increases, which marks the transition from doubly to singly ionised state. The appearance of $U$ band maximum before and $V$ and $R$ band maxima after the $B$ band maximum is consistent with the model of an expanding cooling atmosphere. However, the appearance of $I$ band maximum before $B$ band is in sharp contrast to the simple thermal model \citep{2000A&A...359..876C}.
 
\begin{figure}
\centering
\includegraphics[width=\columnwidth]{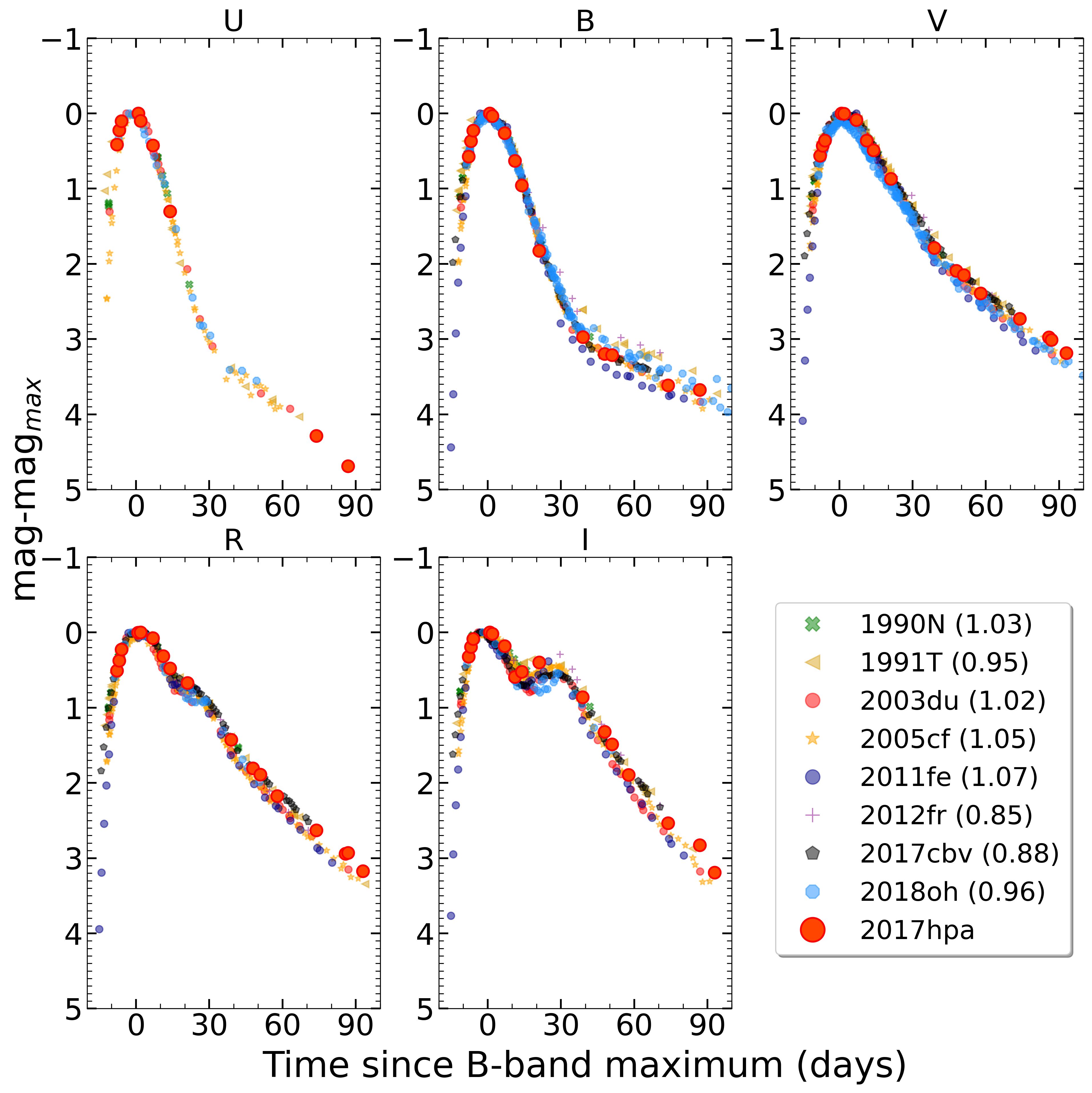}
\caption{$UBVRI$ light curves of SN~2017hpa compared with other normal SN Ia. The light curves have been shifted to match with their peak magnitudes and to the epoch of $B$ maximum.}
\label{Fig3}
\end{figure}

 In Fig.~\ref{Fig3}, the {\it UBVRI} light curves of SN~2017hpa have been compared with other well studied SN Ia, like SN~1990N, SN~1991T \citep{1998AJ....115..234L}, SN~2003du \citep{2005A&A...429..667A, 2007A&A...469..645S}, SN~2005cf \citep{2009ApJ...697..380W}, SN~2011fe \citep{2012A&A...546A..12V}, SN~2012fr \citep{2014AJ....148....1Z}, SN~2017cbv (\citealt{2017ApJ...845L..11H, 2018ApJ...863...90W}) and SN~2018oh \citep{2019ApJ...870...12L}. All the light curves have been shifted to match with their respective $B$ band maximum and to their peak magnitudes. It is evident from Fig.~\ref{Fig3} that the light curves of SN~2017hpa are similar to normal SNe Ia. The decline rate of the light curves has been calculated by a least-square fit to the data around 30--80 days after $B$ band maximum, and the decline rate is given in units of mag (100 d)$^{-1}$. The decline rate in $B$ and $I$ bands are estimated as $2.03\pm0.01$ and $4.48\pm0.01$, respectively.
 
\begin{figure}
\centering
\includegraphics[width=\columnwidth]{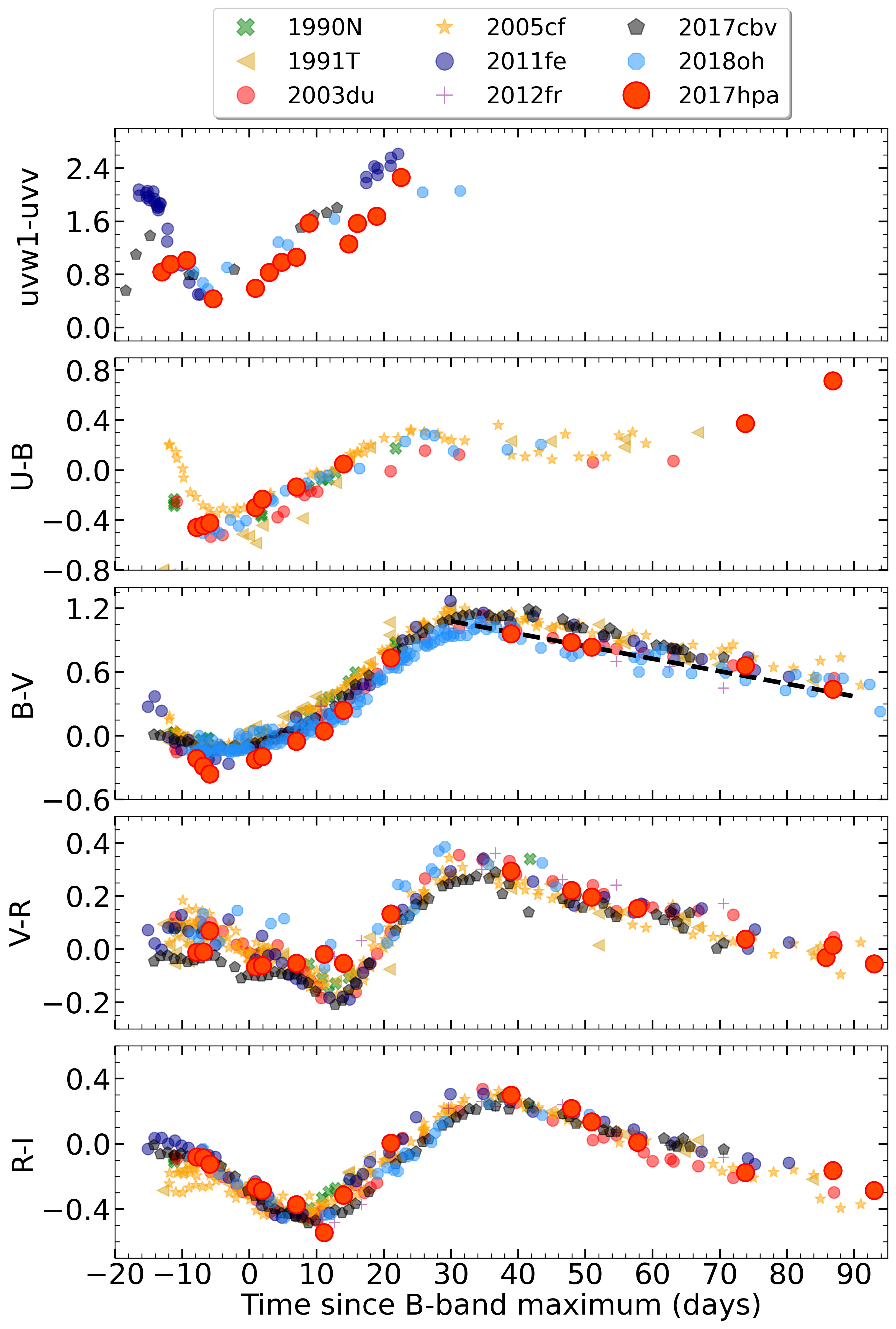}
\caption{$uvw1-uvv$, $U-B$, $B-V$, $V-R$ and $R-I$ colour curves of SN 2017hpa plotted along with other well studied SNe Ia. Plotted as black dotted line in the ($B-V$) colour panel is the Lira relation which has been used to get an estimate of host $E(B-V)$.}
\label{Fig4}
\end{figure}

The reddening corrected $(uvw1-uvv)$, $(U-B)$, $(B-V)$, $(V-R)$ and $(R-I)$ colour curves of SN~2017hpa is displayed in Fig.~\ref{Fig4} and compared with well-studied events. All the SNe have been corrected for reddening due to Milky Way and host, as mentioned in the respective studies. The reddening correction for SN 2017hpa is discussed in Section \ref{Extinction}. The $(B-V)$ colour at $B$-band maximum is $-0.26\pm0.03$~mag, which is bluer than the comparison SNe. The $(uvw1-uvv)$ colour of SN~2017hpa is bluer, similar to the other NUV-blue objects (see Fig.~\ref{Fig4}) and hence can be included in the NUV-blue group as defined by \citet{2013ApJ...779...23M}. However, the $(uvw2-uvv)$ colour evolution could not be verified. Recent studies of carbon positive SNe have shown bluer near UV colours \citep{2011ApJ...743...27T, 2012MNRAS.425.1917S, 2013ApJ...779...23M}, which could be due to unburned carbon present during pre-maximum phases. The $(U-B)$ colour at maximum is slightly redder owing to the absorption by IGE's at shorter wavelengths. This kind of trend has also been seen in SN~2018oh \citep{2019ApJ...870...12L}. The $(V-R)$ colour at maximum is $-0.02\pm 0.01$ mag. The $(V-R)$ colour is redder than the comparison SNe around 10 to 20 d past maximum, and it follows the same trend as other SNe in the late phase. The $(R-I)$ colour at maximum is $-0.24\pm0.01$ mag. The main photometric parameters for SN~2017hpa are listed in Table~\ref{Photometric_parameters}.

\begin{table*}
\centering
\label{tab:Photometric_parameters}
\caption{Photometric parameters of SN 2017hpa.}
\begin{tabular}{lcccccc}
\hline\hline
Filter & JD (Max)  & $m_{\lambda}^\text{max}$& $M_{\lambda}^\text{max}$&  $\Delta m_{15}(\lambda)$ &Decline rate\rlap{*} & Colours at $B$ max\\
     &   & & & & (30--80 d) & \\
\hline\hline  
$U$ & 2458064.26   & 15.11 $\pm$ 0.13  & $-$19.97 $\pm$ 0.18 & 1.26 $\pm$ 0.18 & 2.332 $\pm $0.005  & --\\
$B$ & 2458066.29   & 15.48 $\pm$ 0.11  & $-$19.45 $\pm$ 0.15 & 0.98 $\pm$ 0.16 & 2.026 $\pm$ 0.007 & --\\
$V$ & 2458067.31   & 15.35 $\pm$ 0.09  & $-$19.34 $\pm$ 0.13 & 0.59 $\pm$ 0.13 & 3.155 $\pm$ 0.005 & $(B-V)_0$ = $-$0.26 $\pm$ 0.03 \\
$R$ & 2458067.31   & 15.24 $\pm$ 0.08  & $-$19.31 $\pm$ 0.13 & 0.67 $\pm$ 0.12 & 3.605 $\pm$ 0.004 & $(V-R)_0$ = $-$0.02 $\pm$ 0.01 \\
$I$ & 2458063.25   & 15.36 $\pm$ 0.07  & $-$19.05 $\pm$ 0.13 & 0.62 $\pm$ 0.12 & 4.483$ \pm$ 0.005 & $(R-I)_0$ = $-$0.24 $\pm$ 0.01\\
\hline
\multicolumn{6}{l}{\rlap{*}\  ~in unit of mag (100 d)$^{-1}$ and epoch is relative to $B$ band maximum.}
\end{tabular}
\label{Photometric_parameters}
\end{table*}  

\subsection{Extinction and Distance Modulus}
    The dust map of \cite{2011ApJ...737..103S} gives a colour excess of $E(B-V)$= $0.1518\pm0.0069$ along the direction of SN~2017hpa due to the ISM within the Milky Way. The reddening of the supernova within the host galaxy can be estimated using various empirical relations.
    
    \citet{2003fthp.conf..200T} and \citet{2012poznanski} have shown that the reddening within the host is correlated with the equivalent width of Na\,{\sc i} D. We do not detect Na\,{\sc i} D absorption feature in our low-resolution spectra near maximum light at the redshift of the host galaxy. This indicates minimal reddening within the host. The Phillips-Lira relation \citep{1999AJ....118.1766P}, which uses the $(B-V)$ colour evolution during 30 to 90 days with respect to the $V$-band maximum, can be used to estimate host reddening. After correcting the $(B-V)$ colour curve for reddening in the Milky-Way, the $(B-V)$ colour curve of SN~2017hpa matches well with the colour predicted by Lira relation if it is shifted by a value of $0.08\,\pm\,0.06$ mag. After including the intrinsic dispersion of the Lira relation of $\sim$0.05 mag in quadrature, the reddening estimated from the tail of the $(B-V)$ colour evolution $E(B-V)_{tail}$ is $0.08\pm0.08$ mag. Note the large scatter in the value, which is due to the difference in the slope of the observed $(B-V)$ with the Lira relation. \citet{1999AJ....118.1766P} also used $B_{\rm{max}}$--$V_{\rm{max}}$ color index to estimate $E(B-V)_{\rm{host}}$. We estimate $E(B-V)_{\rm{max}}$ to be $0.06\pm 0.05$ mag. We take an average of $E(B-V)_{\rm{tail}}$ and $E(B-V)_{\rm{max}}$ to get an estimate of $E(B-V)_{\rm{host}}$ to be $0.07\pm0.09$ mag using the $(B-V)$ colours, a value not significantly different from zero.
    
    We have used light curve fitting method MLCS2k2\footnote{\url{https://www.physics.rutgers.edu/~saurabh/mlcs2k2/}} to estimate the host reddening and distance modulus, and SALT2 to confirm it. We applied the updated version of MLCS2k2 (\citealt{1995ApJ...438L..17R,2007ApJ...659..122J}) in which the calibration was improved by applying on a sample on 133 SNe Ia for training and extending the model by incorporating $U$ band data. In this version, the observed LC of a SN Ia in each passband $X$ can be expressed as follows,
    
    \begin{equation}
    \vec{m_{X}}(t-t_{0}) = \vec{M_{X}^{0}} + \mu_{0} + \vec{\zeta_{X}} (\alpha_{X} + \frac{\beta_{X}}{R_{V}})A_{V}^{0} + \vec{P_{X}}\Delta + \vec{Q_{X}}\Delta^{2},
    \end{equation}
    
    where $t_{0}$ is the epoch of maximum light in $B$-band, $\vec{M_{X}^{0}}$ is the absolute magnitudes of the fiducial SN, $\mu_{0}$ is the true distance modulus, $R_{V}$ and $A_{V}^{0}$ are the host galaxy extinction parameters, $\Delta$ is the luminosity/light curve shape parameter, and $\vec{P_{X}}$ and $\vec{Q_{X}}$ are the vectors describing the change in light curve shape as a quadratic function of $\Delta$. The arrowed quantities span the SN rest frame phase. 
    
\begin{figure*}
  \centering
  \begin{subfigure}[b]{0.49\textwidth}
  \resizebox{\linewidth}{!}{\includegraphics{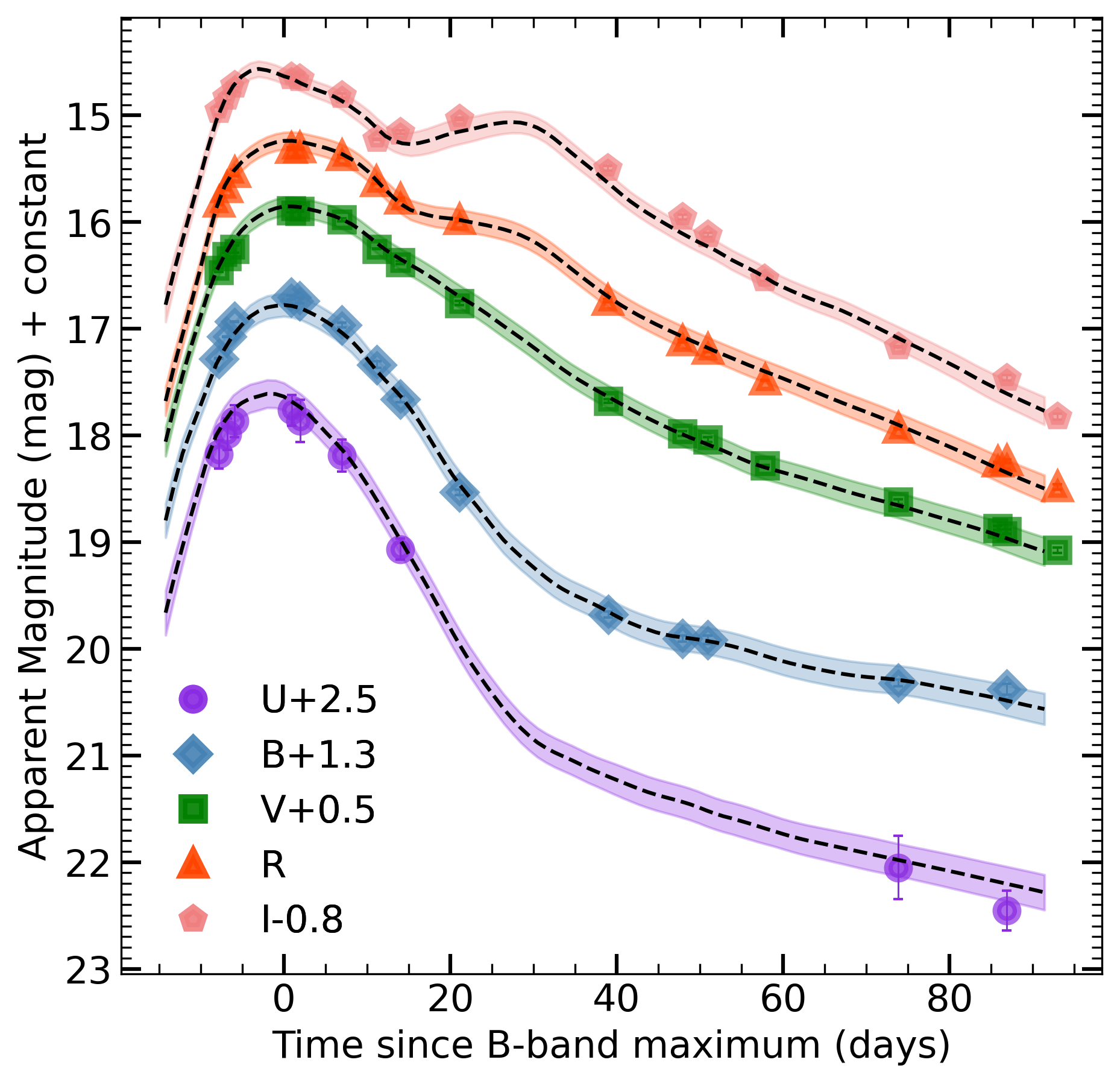}}
  \caption{MLCS2k2 fit}
  \label{Fig5a}
  \end{subfigure}
  \quad
  \begin{subfigure}[b]{0.49\textwidth}
  \resizebox{\linewidth}{!}{\includegraphics{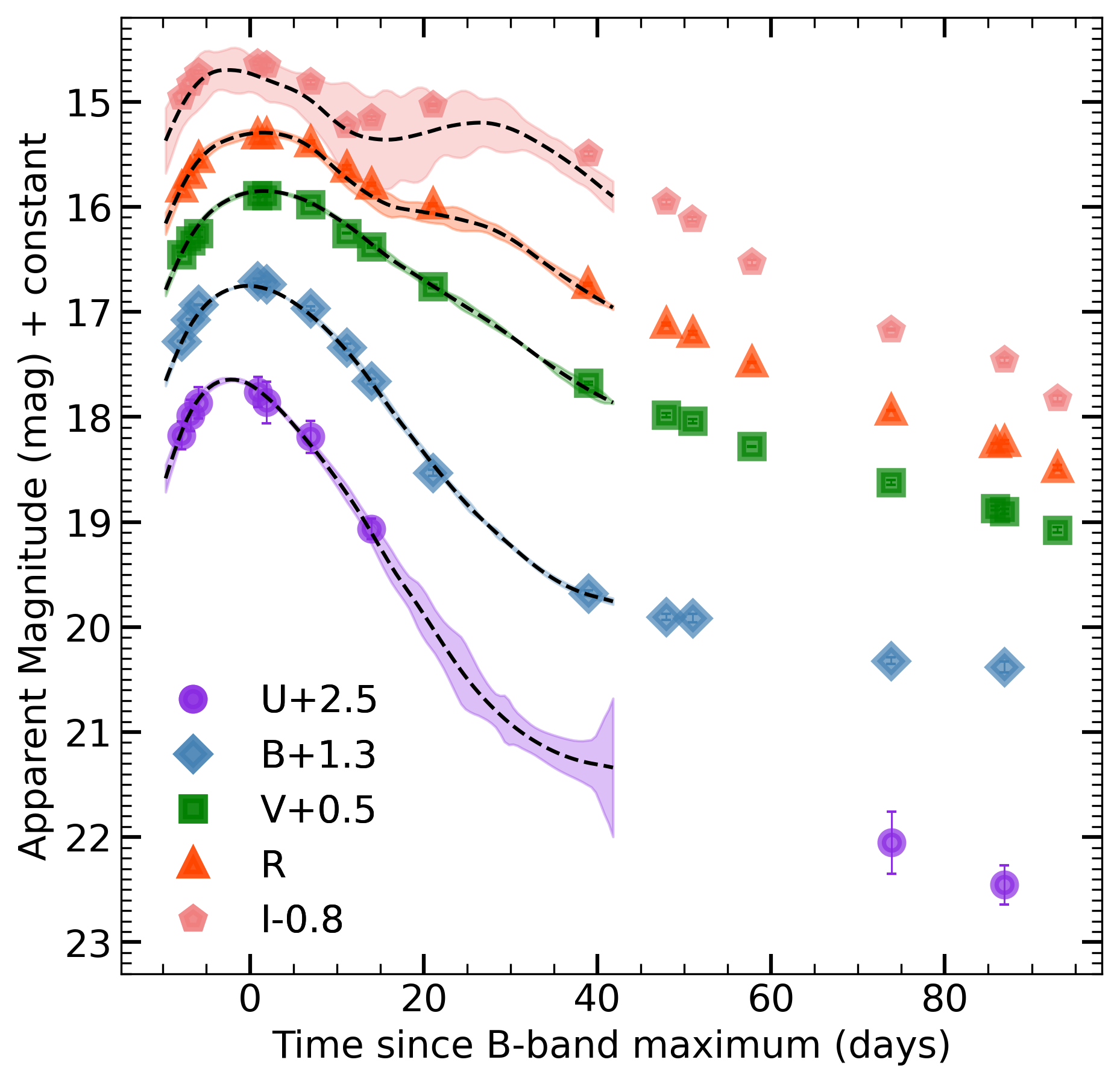}}
  \caption{SALT2 fit}
  \label{Fig5b}
  \end{subfigure}
\caption{Best fit template light curves to the observed $UBVRI$ light curves of SN 2017hpa.}
\label{Fig5}
\end{figure*}

    We fit our observed $UBVRI$ data with MLCS2k2 and kept the ratio of total-to-selective extinction $R_{V}$ to be fixed at the Galactic value of 3.1. With this value of $R_{V}$ we get an extinction $A_{V}^{0}$ of $0.25\pm 0.09$ and a distance modulus $\mu$ of $33.99\pm 0.06$. In their study of 185 SNe Ia in the CfA3 sample, \citet{2009ApJ...700..331H} indicate that the Galactic $R_{V}$ leads to an overestimate of the host extinction, while $R_{V}$ of 1.7 based on their MLCS2k2 fits provides a more realistic extinction estimate. Recent studies have also indicated a non-Milky Way extinction law for the hosts of SNe Ia. For example, an $R_{V}$ of 1.7 was estimated for SN~2017cfd \citep{2020ApJ...892..142H}, and an $R_{V}$ of 2.0 was used for the host of SN~2017cbv and SN~2013aa \citep{2020ApJ...895..118B}. Hence, we proceeded with the fit again, keeping $R_{V}$ as a fit parameter, which resulted in a lower value of $R_{V}= 1.9$ for the host galaxy and the lowest \(\chi^{2}\) for the fit. The change in the value of $R_{V}$ has a negligible effect on the distance when the extinction is low because, for such low reddening it is not possible to separate the effect of $R_{V}$ on the colour of the SN from the intrinsic colour differences between the observed SN and the template. Fig.~\ref{Fig5a} shows the best fit MLCS2k2 templates to the observed data. With this fit we estimate a distance modulus of $34.07\pm 0.06$ mag, and using \cite{1999PASP..111...63F} extinction law, the absolute magnitude in $B$-band is $-19.45\pm 0.15$~mag. The estimate of $E(B-V)=0.08$ matches well with the value estimated based on the observed $(B-V)$ colours at maximum.
    
    We further fit the light curve with Spectral Adaptive Light Curve template (SALT2\footnote{\url{http://supernovae.in2p3.fr/salt/doku.php}}), which models the spectral energy distribution of SN Ia as:
    
    \begin{equation}
    F_{\rm{SN}}(p, \lambda) = x_{0}(M_{0}(p, \lambda)+x_{1}M_{1}(p, \lambda))\exp(cC_{L}(\lambda)),
    \end{equation}
    
    where, $F_{\rm{SN}}$ is the phase-dependent flux density in the rest-frame of the SN, $p$ = t -- \(\rm t_{0}\) is the phase of the SN, $x_{0}$, $x_{1}$, and $c$ are the normalisation, shape, and colour parameter respectively. $M_{0}$, $M_{1}$ and $C_{L}$ are the mean spectral sequence, first-order deviation around the mean sequence and time-independent colour law. These are the trained vectors of SALT2. We applied the SALT2 model \citep{2014A&A...568A..22B} with the Landolt-Bessell filter set and used the Vega magnitude system. We used only the observed $UBVRI$ data in the fit. The best-fit templates to the observed data are shown in Fig. \ref{Fig5b}.
    
    SALT2 does not fit the distance as a parameter. We therefore calculated the distance modulus using the following relation:
    
    \begin{equation}
    \mu_{0} = m_{B}^{*} - M_{B} + \alpha x_{1} - \beta c,
    \end{equation}
    
    where, $m_{B}^{*}$, $x_{1}$ and $c$ are the fit parameters from SALT2 and $M_{B}$, $\alpha$ and $\beta$ are parameters for the distance estimate. We used the values of $M_{B}$, $\alpha$ and $\beta$ from \citet{2014A&A...568A..22B}. The $M_{B}$ parameter used in the calibration is valid for SNe which exploded in galaxies having total stellar mass $M_{\rm{stellar}} \le 10^{10}\, M_\odot$. For a more massive host galaxy, a correction of -0.061 has to be added to $M_{B}$.

    As the value of $H_{0}$ adopted in MLCS2k2 and SALT2 is 65 km s$^{-1}$Mpc$^{-1}$ and 68 km s$^{-1}$Mpc$^{-1}$, respectively we have converted them to $H_{0} = 73$ km s$^{-1}$Mpc$^{-1}$ by using the following equation: 
    
    \begin{equation}
    \mu_{0}(H_{0}) = \mu_{0}(\rm{Model}) - 5log_{10}(H_{0}/H_{0}^{Model})~mag,
    \end{equation}
    
    where "Model" refers to MLCS2k2/SALT2. From SALT2 fit we get a distance modulus of 34.09\,$\pm$\,0.08 mag. The distance modulus is 34.15\,$\pm$\,0.08 for a host galaxy of stellar mass $\ge 10^{10}\, M_\odot$. The main parameters of both the model fits are listed in Table~\ref{tab:LC_Fit}.

Throughout the entire analysis, we have used an $R_{V}=3.1$ for the Milky Way and an $R_{V}=1.9$ for the host galaxy obtained from MLCS2k2 fit. The $E(B-V)$ used for the host is $0.08\pm0.06$ mag. The distance modulus of $34.08\pm0.09$ mag of the SN is obtained by averaging the values from SALT2 and MLCS2k2 fits, assuming the host galaxy stellar mass is $\le 10^{10}\, M_\odot$. 

\label{Extinction}

\begin{table}
\caption{Best-fit parameters of the light curve for 2017hpa.}
\label{tab:LC_Fit}
\centering
\begin{tabular}{lcc}
\hline
\hline
%\noalign{\smallskip}
Parameter & Value & Error \\
\hline
MLCS2k2 & \multicolumn{2}{c}{\(UBVRI\)}\\
\hline
R\(_{V}\) & 1.9 &  \\
T\(_{max}\) (MJD) & 58066.29 & 0.11 \\
A\(_{V}^{host}\) (mag) &  0.16 & 0.06 \\
\(\Delta\) (mag) & --0.17 & 0.03 \\
\(\mu_{0}\) (H\(_{0}\)=73) (mag)  & 34.07 & 0.05 \\
\(\chi^{2}\)/dof &  0.78 \\
\hline
\hline
\end{tabular}
\quad
\centering
\begin{tabular}{lcc}
\hline
\hline
%\noalign{\smallskip}
Parameter & Value & Error \\
\hline
SALT2.4 & \multicolumn{2}{c}{\(UBVRI\)}\\
\hline
T\(_{max}\) (MJD) & 58066.89 & 0.06 \\
\(c\) & --0.07 & 0.03 \\
x\(_{0}\) & 0.02 & 0.00 \\
x\(_{1}\) & 0.19 & 0.07 \\
m\(_{B}^{*}\) (mag) & 14.84 & 0.03 \\
\(\mu_{0}\) (H\(_{0}\)=73) (mag) & 34.09 & 0.08 \\
\(\chi^{2}\)/dof  &  2.67 \\
\hline
\hline
\end{tabular}
\end{table} 

\subsection{Estimation of Nickel Mass}
    
\begin{figure}
\centering
\resizebox{\hsize}{!}{\includegraphics[width=\columnwidth]{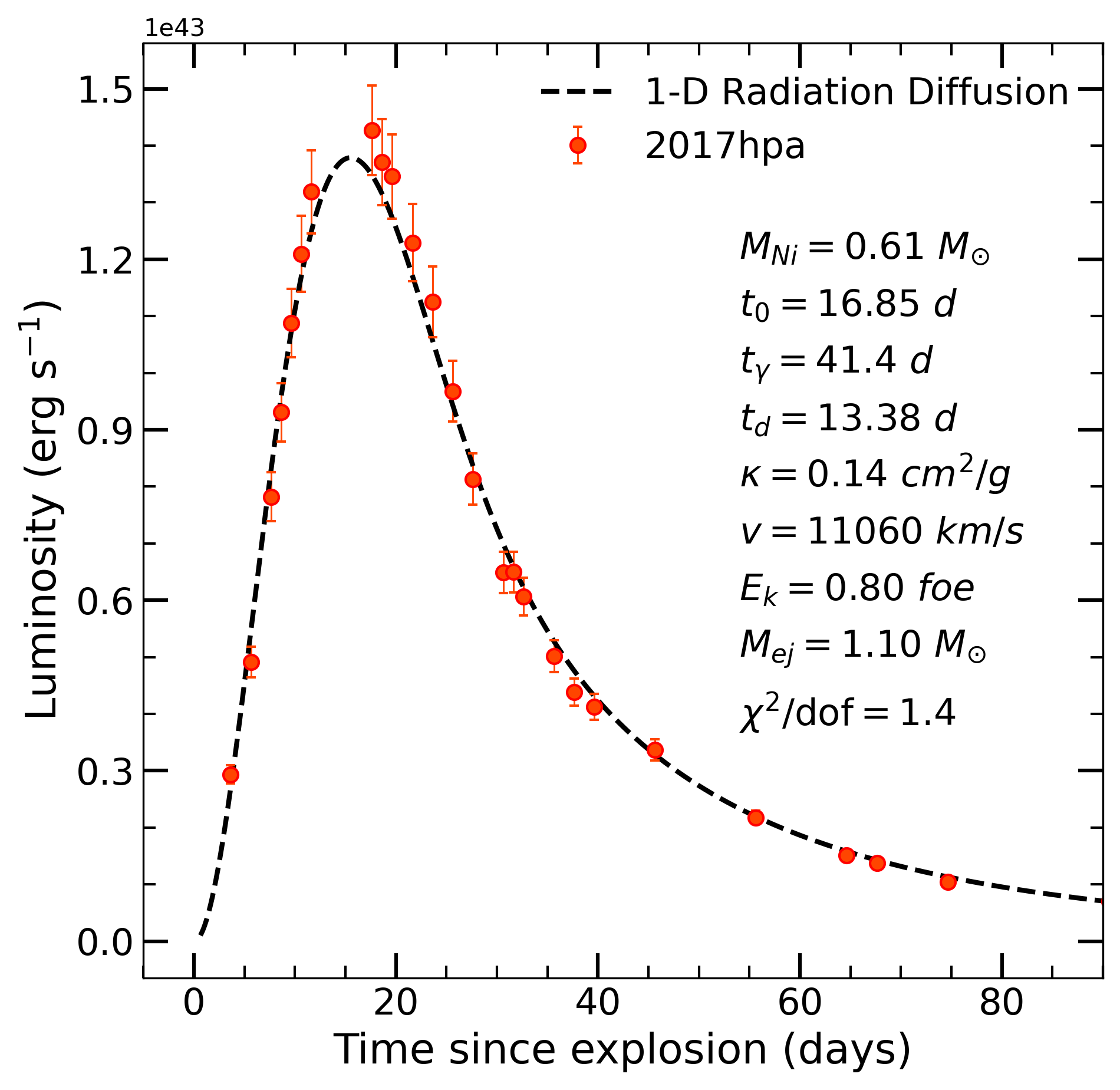}}
\caption{The bolometric light curve of SN~2017hpa plotted along with the 1-D Radiation Diffusion model fit.}
\label{Fig6}
\end{figure}

    The bolometric light curve, constructed using the $U$, $B$, $V$, $R$, $I$ and $uvw1$ magnitudes, is shown in Fig.~\ref{Fig6}. The apparent magnitudes were corrected for Milky-Way reddening of $E(B-V)$ \(=\) 0.1518 \(\pm\) 0.0069 with R\(_{V}\) = 3.1 and a host galaxy reddening of $E(B-V)$ \(=\) 0.08 $\pm$ 0.06 with R\(_{V}\) = 1.9. The reddening corrected magnitudes were converted to flux using zero-points from  \cite{1998A&A...333..231B}. Magnitudes in $U$, $B$, $V$, $R$ and $I$ have been interpolated from MLCS2k2 model light curve and Swift-$UVOT$ band has been interpolated by fitting a  cubic spline. Then, the spectral energy distribution (SED) is constructed by using the fluxes in $UBVRI$ and $UVOT$. A spline curve has been fit to the SED and the area under the curve has been calculated by trapezoidal rule, integrating from 2500--9500 \text{\AA}. The NIR (9500-24000 \text{\AA}) flux contribution has been estimated from \citet{2009ApJ...697..380W} and added to the uv-optical flux. The total flux thus obtained has been converted to luminosity adopting a distance modulus of 34.08. The maximum of the bolometric light curve for SN~2017hpa is $L^{\rm{bol}}_{\rm{peak}}=1.43\times10^{43}$ erg s$^{-1}$. To estimate nickel mass, ejecta mass and other physical parameters of the explosion, we applied the modified radiation diffusion model (\citealt{1982ApJ...253..785A, 2008MNRAS.383.1485V, 2012ApJ...746..121C}). The model assumes homologously expanding ejecta, spherical symmetric distribution of ejecta material, no mixing, constant optical opacity, small initial radius (R\(_{0}\) \(\sim\)\,0) and the diffusion approximation for photons. To fit the model we used a Markov-Chain Monte Carlo method. We used the {\small \it{emcee}} package in {\small \textit{python}} and optimized the parameters and sampled the posterior distribution following the methods as described in \citet{2013PASP..125..306F}. The fit parameters of the model are $t_0$ - the rise time to maximum since the day of explosion, $M_{\rm{Ni}}$ - the initial $^{56}$Ni mass produced, $t_{\rm{lc}}$ -  the light curve time scale and $t_\gamma$ - the gamma-ray leaking time scale. The fit to the bolometric light curve gives a rise time to maximum $t_{0}=16.93\pm 0.23$ days, $t_{\rm{lc}}=13.38\pm0.47$ days, $t_\gamma=41.4\pm 0.8$ days and $M_{\rm{Ni}}=0.61\pm0.02\,M_\odot$. 
    
    The ejected mass ($M_{\rm{ej}}$) and the expansion velocity ($v_{\rm{exp}}$) are related to the two timescale parameters $t_{\rm{lc}}$ and $t_{\gamma}$ of the radiation model by
    
    \begin{equation}
    t_{\rm{lc}}^{2} = \frac{2\kappa M_{\rm{ej}}}{\beta c v_{\rm{exp}}} \ \text{and} \ t_{\gamma}^{2} = \frac{3\kappa_{\gamma}M_{\rm{ej}}}{4\pi v_{\rm{exp}}^{2}}.
    \end{equation}
    
    Here, $\kappa$ is the effective optical opacity, $ \kappa_{\gamma}=0.03$~cm\(^{2}\) g\(^{-1}\) is the opacity for $ \gamma$-rays (\citealt{1997ApJ...491..375C, 2015MNRAS.450.1295W}), and $\beta=13.8$ is a constant of integration \cite{1982ApJ...253..785A}. These two equations can be used to constrain three parameters of explosion, $M_{\rm{ej}}$, $v_{\rm{exp}}$ and $\kappa$. Following \citep{2019ApJ...870...12L} and \citep{2020ApJ...892..121K}, we can get a lower bound on $\kappa$ assuming $M_{\rm{ej}} \le  M_{\rm{Ch}}$, while the upper limit on $\kappa$ is obtained assuming $v_{\rm{exp}}$ has a lower limit of 9500 km s\(^{-1}\), estimated from the Si\,{\sc ii} \(\lambda\)6355 absorption feature near maximum light. In this way, the limits of $\kappa$, i.e.\ $\kappa_{lower} = 0.12 \pm 0.01$~cm\(^{2}\) g\(^{-1}\) and $\kappa_{upper} = 0.16 \pm 0.01$~cm\(^{2}\) g\(^{-1}\), are estimated, and the effective optical opacity is taken to be $\kappa = 0.14 \pm 0.01$. Finally, by using the equations
    
    \begin{equation}
    M_{\rm{ej}} = \frac{3\kappa_{\gamma}t_{lc}^{4}\beta^{2}c^{2}}{16\pi t_{\gamma}^{2}\kappa^{2}}; \ v_{\rm{exp}}  = \frac{3\kappa_{\gamma}t_{\rm{lc}}^{2}\beta c}{8\pi \kappa t_{\gamma}^{2}} \ {\rm{and}} \ E_{{kinetic}} = 0.3M_{{ej}}v_{{exp}}^{2},
    \end{equation}
    
    we estimate $M_{\rm{ej}} = 1.10 \pm 0.22\,M_\odot$, $v_{\rm{exp}} = 11,060 \pm 1200$ km s$^{-1}$ and kinetic energy $ E_{\rm{kinetic}} = 0.80\pm0.23 \times 10^{51}$ erg.
    
\section{Spectral Analysis}
\label{Spectroscopy}

The spectra of SN~2017hpa were obtained from days $-8.9$ to $+108$ since the $B$-band maximum. In our analysis, we have also used the earliest spectrum available for SN~2017hpa in WISeREP \citep{2012PASP..124..668Y},\footnote{\url{https://wiserep.weizmann.ac.il/}} taken at $-13.8$~d with the Asiago Faint Object Spectrograph and Camera (AFOSC). All spectra were corrected for a redshift of 0.015 and a reddening value of $E(B-V)=0.15$ mag for the Milky Way and $E(B-V)=0.08$ mag for the host galaxy. The spectroscopic features have been identified by comparing with other SNe Ia around similar phases.

\subsection{Pre-maximum  phase}
\begin{figure}
\centering
\resizebox{\hsize}{!}{\includegraphics[width=\columnwidth]{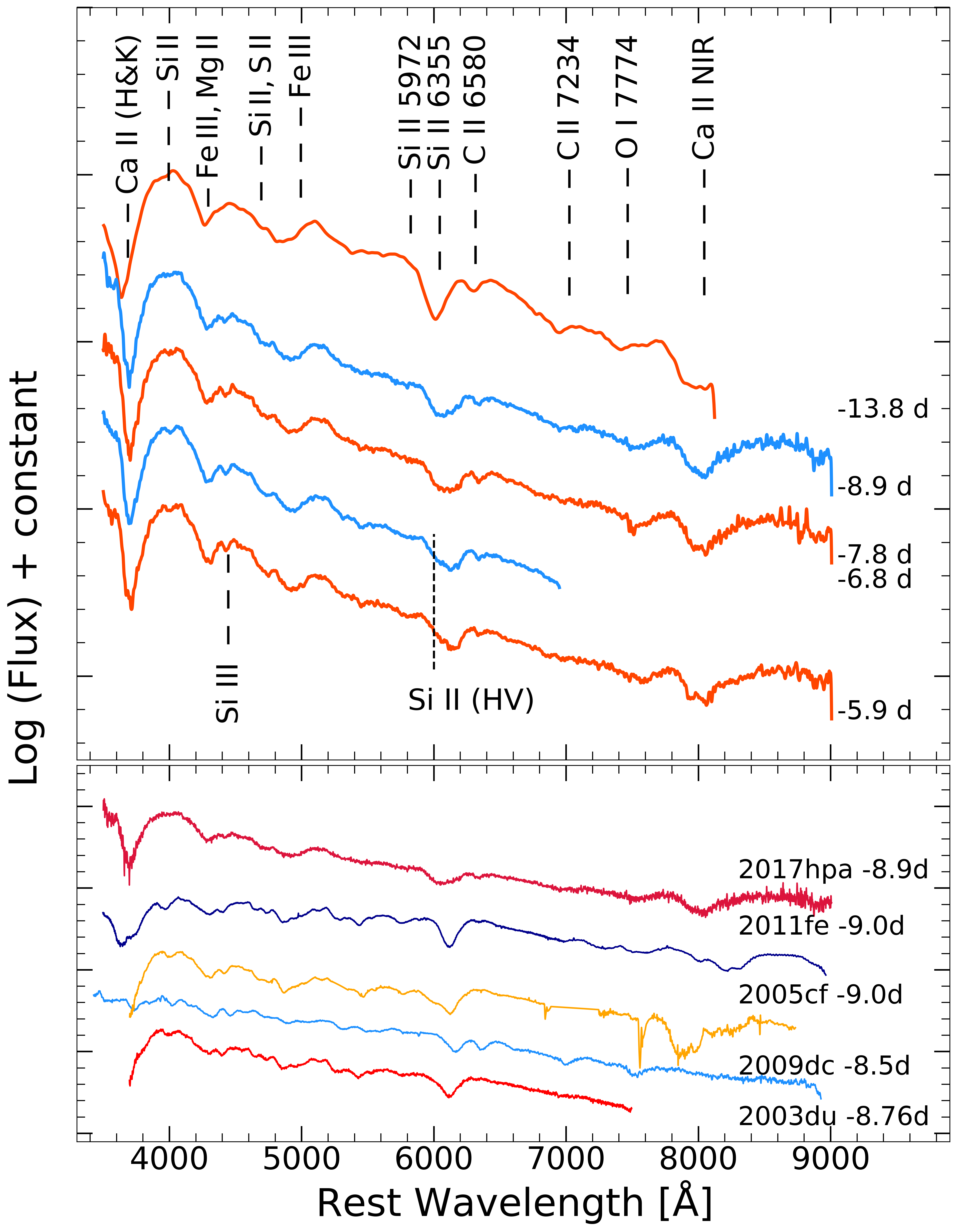}}
\caption{Top panel: Evolution of the pre-maximum spectrum of SN~2017hpa. The $-13.8$ d spectrum downloaded from WISeReP is also plotted. Bottom panel: The $-8.9$ d spectrum of SN~2017hpa has been compared with SN~2011fe, SN~2009dc, SN~2005cf and SN~2003du around similar phase. The spectra have been smoothed while plotting to enhance visibility of the features.}
\label{Fig7}
\end{figure}
 
The pre-maximum spectra during $-13.8$ to $-5.9$ d are plotted in Fig~\ref{Fig7}. The spectra show blue continuum with prominent absorption due to {Ca\,{\sc ii} H \& K ($\lambda\lambda$3934, 3968)}, Fe\,{\sc iii}, Mg\,{\sc ii}, Si\,{\sc iii} and Si\,{\sc ii} ($\lambda$6355). The absorption feature of Ca\,{\sc ii} NIR triplet slowly starts developing. As the SN evolves, the above lines become stronger, and the lines due to the iron group elements start appearing in the spectra. In the pre-maximum phase, the profile of Si\,{\sc ii} ($\lambda$6355) line is broad and asymmetric with respect to line minimum. This is due to the presence of a high-velocity component in the spectra. The high-velocity feature is usually seen in the Ca\,{\sc ii} NIR triplet in a significant fraction of type Ia supernova observed during pre-maximum phase (\citealt{2014MNRAS.444.3258M,2017MNRAS.467..778M}). In the emission wing of the Si\,{\sc ii} $\lambda$6355 line, a prominent absorption feature is seen at $\sim$\,6300 \AA\ which is due to blueshifted C\,{\sc ii} ($\lambda$6580) absorption feature.  This can be explained by the presence of unburned C that is attributed to subsonic burning (deflagration wave) in a significant fraction of the white dwarf \citep{2007ApJ...668.1132R}. 

The spectrum at $-13.8$~d shows weak blended lines of Fe\,{\sc ii} ($\lambda$4924, 5018), Fe\,{\sc iii} ($\lambda$4421, 5075, 5158), Co\,{\sc ii} ($\lambda$4161) Mg\,{\sc ii} ($\lambda$4481, 7890), S\,{\sc ii} ($\lambda$4716, 5032, 5321, 5429, 5454, 5510), Si\,{\sc ii} ($\lambda$4128, 5972, 6355), Si\,{\sc iii} ($\lambda$4568), which become prominent around the maximum. Also, Si\,{\sc ii} ($\lambda$5972) starts appearing at $-6$~d. Strong Si\,{\sc ii} ($\lambda$6355) line and weak Si\,{\sc ii} ($\lambda$5972) indicate that the photosphere is hot. 

The pre-maximum spectrum of SN~2017hpa at $-8.9$~d is compared with those of SN~2003du (\citealt{2012AJ....143..126B}), SN~2005cf (\citealt{2007A&A...471..527G}), SN~2009dc (\citealt{2011MNRAS.412.2735T}), SN~2011fe (\citealt{2016ApJ...820...67Z}). Among these events, SN~2003du, SN~2005cf and SN~2011fe are normal type Ia SNe, whereas SN~2009dc belongs to the super-Chandrasekhar mass category. All the compared spectra (see Fig.~\ref{Fig7} bottom panel) exhibit C\,{\sc ii} ($\lambda$6580) absorption feature during this phase. However, the C\,{\sc ii} ($\lambda$6580) feature is more prominent in SN~2017hpa than normal SNe Ia, similar to SN~2009dc. By fitting a Gaussian to the absorption feature of C\,{\sc ii} $\lambda$6580, we estimate that the pEW of C\,{\sc ii} evolves from $\rm 13.56\pm1.49$ \AA\ at $-13.8$~d to $\rm 4.49\pm0.43$ \AA\ at $-5.9$~d. The Si\,{\sc ii} $\lambda$5972 line is weaker than other objects at this phase.  The IME signatures are weak in SN~2017hpa while these are prominent in other objects (see bottom panel of Fig.~\ref{Fig7}). The relatively weak IME features suggest ongoing burning \citep{2004MNRAS.348..261B, 2011MNRAS.412.2735T}. The line formation not only depends on the composition but also on the ionisation state of the ejecta. The weakness of the singly ionised IME features can also be explained by higher temperature and in turn, higher ionisation. 

\begin{figure}
\centering
\resizebox{\hsize}{!}{\includegraphics[width=\columnwidth]{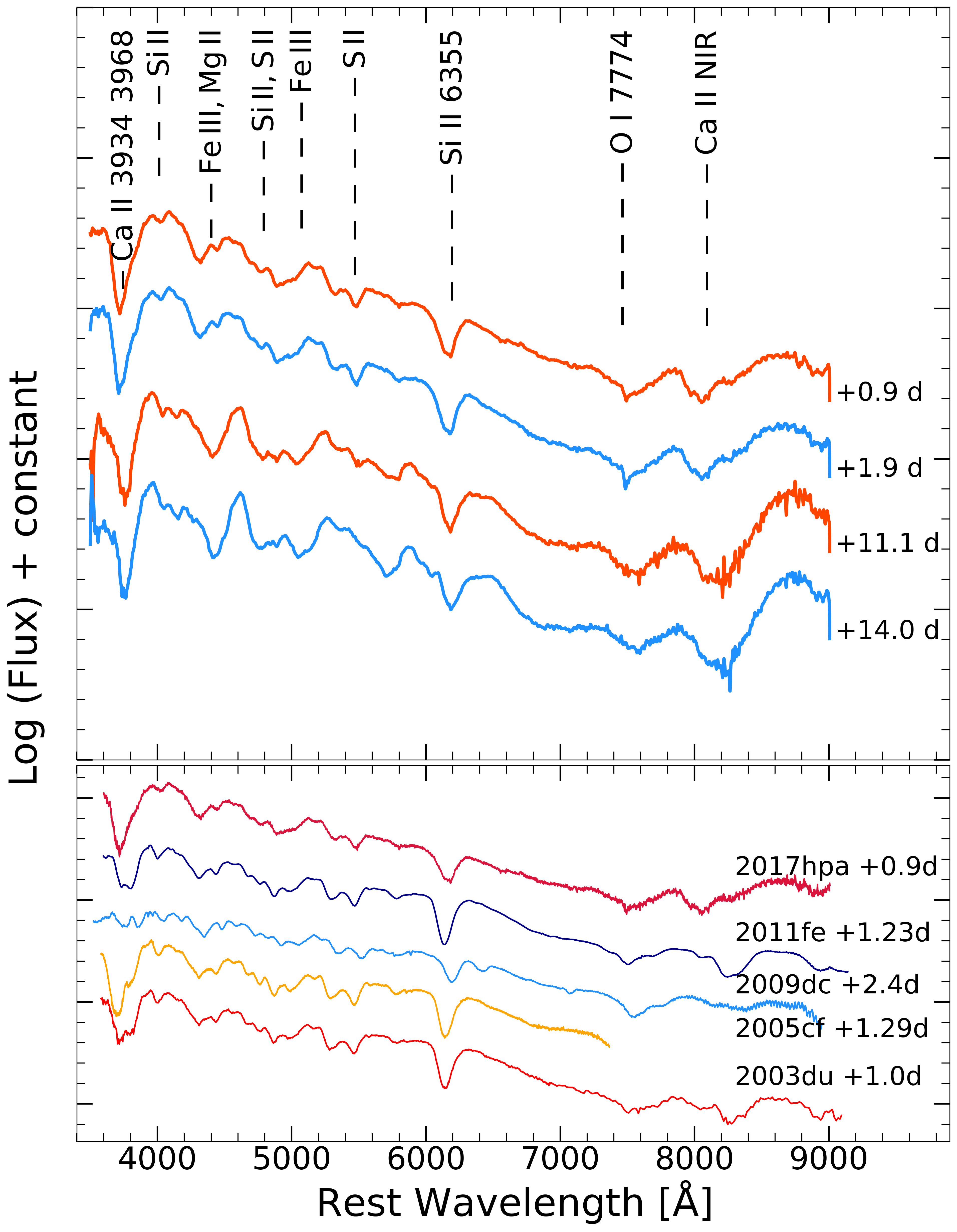}}
\caption{Top panel: Spectral evolution of SN 2017hpa around maximum. Bottom panel: The $+0.9$ d spectrum of SN~2017hpa has been compared with SN~2011fe, SN~2009dc, SN~2005cf and SN~2003du around similar phase. The spectra have been smoothed while plotting to enhance visibility of the features.}
\label{Fig8}
\end{figure}

The flat-bottom, broad profile of Si\,{\sc ii} $\lambda$6355 (at $-13.8$~d, $-8.9$~d and $-7.8$~d) and the asymmetric profile (at $-6.8$~d and $-5.9$~d) is most likely due to a high velocity component detached from the photosphere. A high velocity component in Si\,{\sc ii} has been observed in other well studied type Ia SNe like SN~1990N \citep{2001MNRAS.321..341M}, SN~2003du (\citealt{2005A&A...429..667A, 2011MNRAS.410.1725T}), SN~2005cf \citep{2009ApJ...697..380W}, SN~2009ig (\citealt{2012ApJ...744...38F, 2013ApJ...777...40M}), SN~2012fr \citep{2013ApJ...770...29C}, SN~2019ein (\citealt{2020ApJ...893..143K, 2020ApJ...897..159P}). By fitting a Gaussian to the absorption profile of Si\,{\sc ii} ($\lambda$6355) and Si\,{\sc ii} ($\lambda$5972) at $-5.9$~d we estimate the pEW of Si\,{\sc ii} ($\lambda$5972) to be $\rm 13.75\pm1.70$ \AA\ and that of Si\,{\sc ii} ($\lambda$6355) to be $\rm 67.22\pm4.88$ \AA\ , implying a ratio R(Si\,{\sc ii}) of $\rm 0.20\pm0.03$. At $+0.9$~d the pEW of Si\,{\sc ii} ($\lambda$5972) is $\rm 9.22\pm1.02$ \AA\ that of Si\,{\sc ii} ($\lambda$6355) is $\rm 75.16\pm1.96$ \AA\, and R(Si\,{\sc ii}) is $\rm 0.13\pm0.02$. From the value of R(Si\,{\sc ii}),  SN~2017hpa can be placed among the Core-Normal (CN) class described by \cite{2006PASP..118..560B}.

\begin{figure}
\centering
\resizebox{\hsize}{!}{\includegraphics[width=\columnwidth]{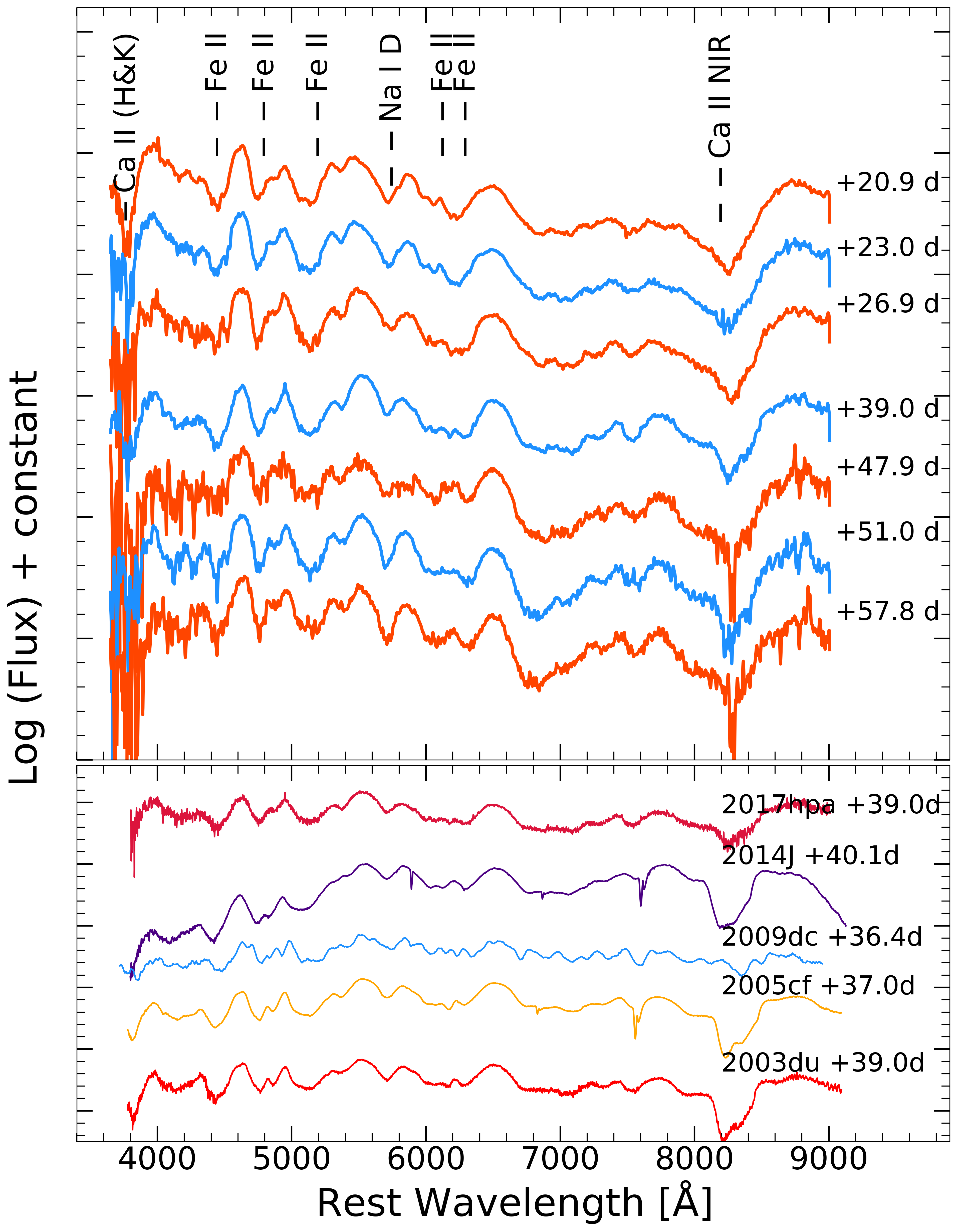}}
\caption{Top panel: Post maximum spectral evolution of SN~2017hpa. Bottom panel: The $+39.0$ d spectrum of SN~2017hpa has been plotted with SN~2014J, SN~2009dc, SN~2005cf, SN~2003du for comparison. The spectra have been smoothed while plotting to enhance visibility of the features.}
\label{Fig9}
\end{figure}

\subsection{Maximum and early post-maximum phase}
The spectral evolution during the maximum and early post-maximum phase is presented in Fig.~\ref{Fig8}. The C\,{\sc ii} ($\lambda$6580) line seen in the pre-maximum phase is not seen in the spectrum taken close to maximum. Around this phase, the IME features have become prominent. The 'W'  feature due to S\,{\sc ii} ($\lambda$5468, 5654) line is prominent in the maximum spectrum, and it becomes weak in the spectrum taken at +11 d. The high-velocity feature in the Si\,{\sc ii} ($\lambda$6355) line seen during the pre-maximum phase disappears close to maximum, and the photospheric component is dominant. Because of the high-velocity feature in the Ca\,{\sc ii}  NIR triplet, the absorption feature shows two minima till +2 d and it is not seen in the next spectrum taken at +11 d. The Si\,{\sc iii} ($\lambda$4568) line was seen in the spectrum till +2 d but was not present in the later spectra. The spectrum of SN~2017hpa close to the maximum is compared with the spectra of some other well studied SNe (see bottom panel of Fig.~\ref{Fig8}) and is found to be similar. 

\subsection{Post-maximum to early nebular phases}
\begin{figure}
\centering
\resizebox{\hsize}{!}{\includegraphics[width=\columnwidth]{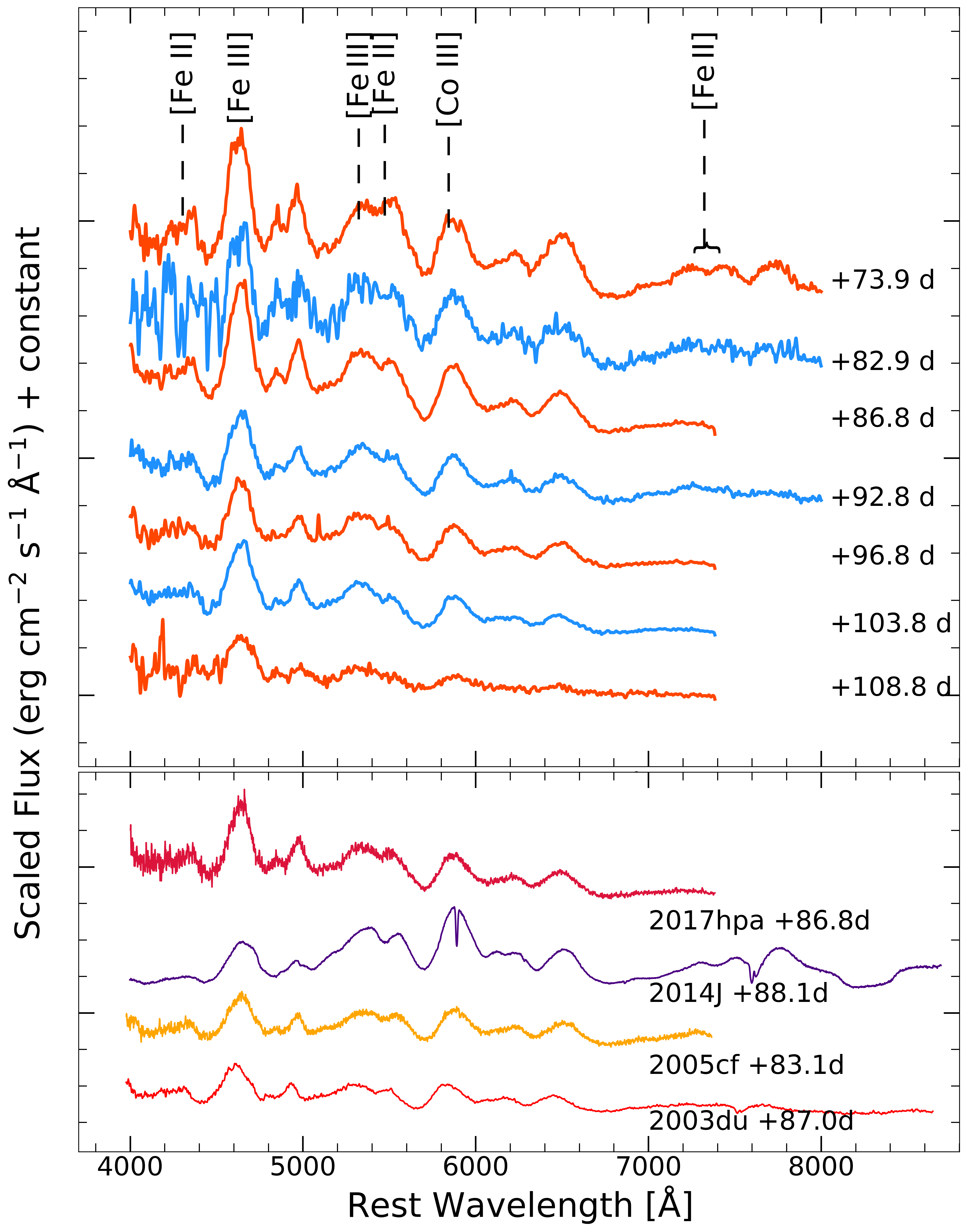}}
\caption{Top panel: Spectral evolution of SN~2017hpa in the early nebular phase. Bottom panel: The $+86.8$ d spectrum of SN~2017hpa has been compared with SN~2014J, SN~2005cf and SN~2003du around similar phases for comparison. The spectra have been smoothed while plotting to enhance visibility of the features.}
\label{Fig10}
\end{figure}

The spectral evolution of SN~2017hpa during post-maximum to early nebular phases is presented in Fig.~\ref{Fig9} and Fig.~\ref{Fig10}.
It is evident from Fig.~\ref{Fig9} that the Si\,{\sc ii} line is weakening, and Fe\,{\sc ii} ($\lambda$6238, 6248) lines are getting stronger. The pEW of Si\,{\sc ii}($\lambda$6355) evolves from $\rm 75.16\pm1.96$ at +0.9 d to $\rm 57.04\pm1.38$ at +14.0 d. This weakening of the Si\,{\sc ii} feature indicates that the ejecta is diluting and the SN enters into the Fe\,{\sc ii} dominated phase. Further, the Ca\,{\sc ii} H \& K feature begins to weaken around +23 d. In contrast, the Ca\,{\sc ii} NIR triplet remains a broad, flat bottom absorption profile with a pronounced emission component until +57 d. 

During the late post maximum phase to the early nebular phase, the ejecta begins to get transparent, and the inner regions of the explosion can be probed. The Fe\,{\sc ii} absorption feature at $\lambda$4555 remains prominent, and the Fe\,{\sc ii} lines at $\lambda$4924 and $\lambda$5018 are clearly seen till +58.0 d. 
The Fe\,{\sc ii} line at $\lambda$5169 shows a deep absorption feature at 5100 \text{\AA}, and to the redward wing, the Fe\,{\sc ii} $\lambda$5536 line can be seen as a prominent notch. Around 6000 \text{\AA} there are blended lines of Fe\,{\sc ii} $\lambda$6238, $\lambda$6248 and $\lambda$6451, $\lambda$6456. Fe\,{\sc ii} lines at $\lambda$7308 and $\lambda$7462 are weak. In the bottom panel of Fig.~\ref{Fig9} the +39.0 day spectra of SN~2017hpa has been compared with SN~2014J \citep{2016MNRAS.457.1000S}, SN~2009dc, SN~2005cf, SN~2003du and SN~2002bo around similar phases.

The spectrum at $+74.0$~d shows strong forbidden emission lines due to Fe and Co. This shows that the SN has entered into the nebular phase \citep{2014MNRAS.441.3249D}. Early nebular phase spectra show features of [Fe\,{\sc iii}], [Fe\,{\sc ii}], [Ni\,{\sc ii}], [Co\,{\sc iii}]. The strongest feature seen in the nebular phase spectra is the emission at $\lambda$4700, which is due to a blend of several [Fe\,{\sc iii}] lines with some contribution from [Fe\,{\sc ii}]. The emission line seen around 5100 \text{\AA} in our early nebular phase spectra is due to contributions from [Fe\,{\sc ii}] and [Fe\,{\sc iii}] lines. In the very early nebular phase spectra, the [Co\,{\sc iii}] emission line at around 5900 \text{\AA} is quite strong, but as the supernova ages, the [Co\,{\sc iii}] lines become weaker due to decay of $^{56}$Co to $^{56}$Fe. In the bottom panel of Fig.~\ref{Fig10} the $+86.8$~d spectrum of SN~2017hpa has been compared with SN~2014J, SN~2005cf and SN~2003du. The [Fe\,{\sc ii}] emission feature around 4700 \AA\ at $+86.8$~d is stronger in the spectrum of SN~2017hpa than the comparison SNe. 

\begin{figure*}
\centering
\resizebox{\hsize}{!}{\includegraphics[width=\textwidth]{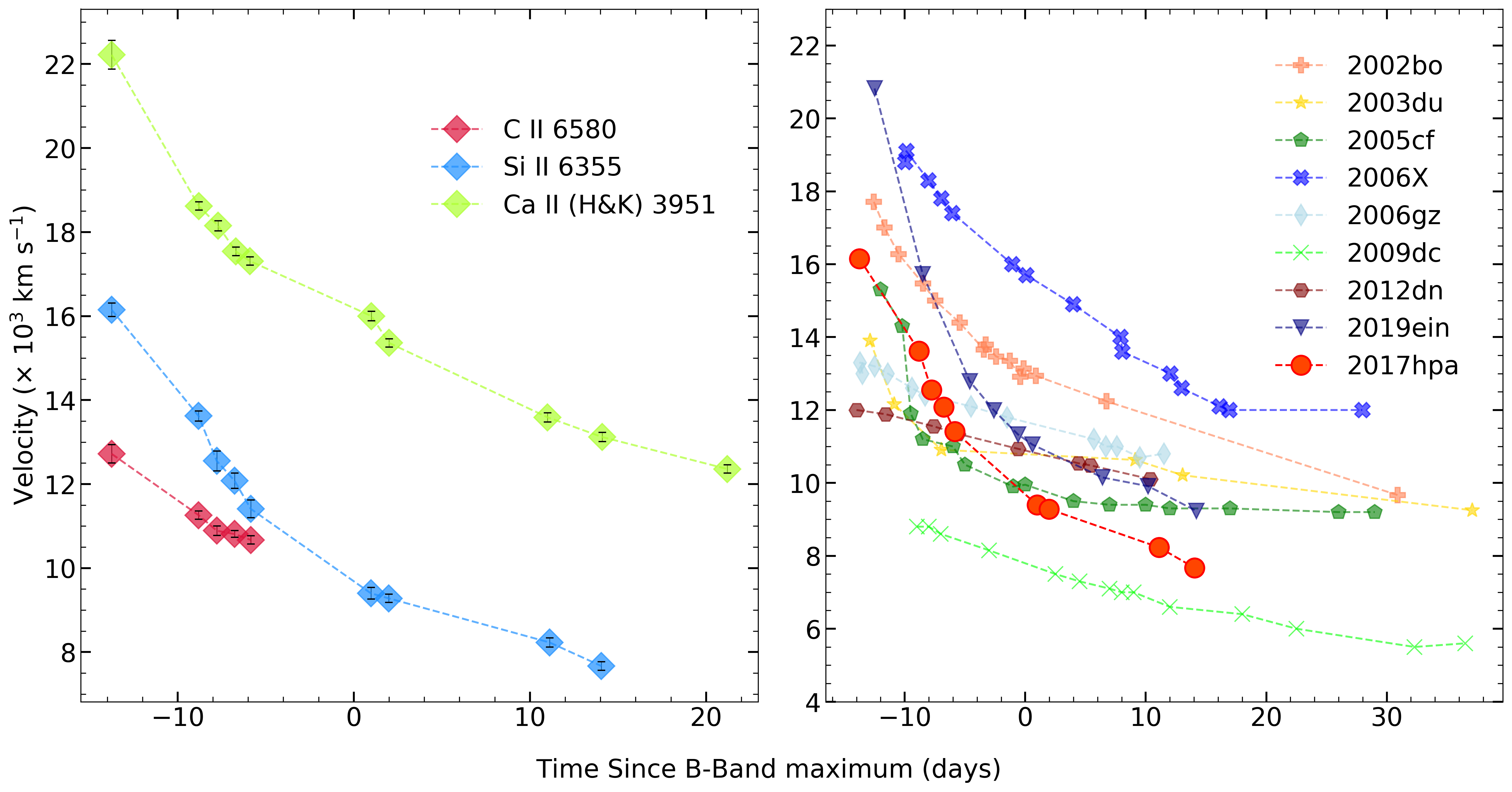}}
\caption{Velocity evolution of C\,{\sc ii} $\lambda$6580, Si\,{\sc ii} $\lambda$6355 and Ca\,{\sc ii} $\lambda$3951 for SN~2017hpa (left panel). Comparison of the velocity evolution of Si\,{\sc ii} $\lambda$6355 for SN~2017hpa with other SNe Ia (right panel).}
\label{Fig11}
\end{figure*} 
    
\subsection{Velocity Evolution}

The velocity of the spectral lines has been measured by fitting a single Gaussian function to the absorption trough of the corresponding lines. The velocities of Ca\,{\sc ii} (H \& K) $\lambda$3951, Si\,{\sc ii} $\lambda$6355, C\,{\sc ii} $\lambda$6580 lines estimated based on the absorption minimum are plotted in Fig.~\ref{Fig11} (left panel). The most prominent line in the spectra, Si\,{\sc ii} $\lambda$6355 traces the evolution of the expansion velocity of the ejecta for over 4--5 weeks since the explosion. The Si\,{\sc ii} line has a velocity of $16,150\pm160$ km s$^{-1}$ around $-13$ d, which is quite normal for type Ia SNe around this phase. The Si\,{\sc ii} line velocity drops rapidly to $9400\pm140$ km s$^{-1}$ close to maximum. The subsequent decline is slow, reaching a velocity of $\sim 8000$ km s$^{-1}$ at +15 d. 

The C\,{\sc ii} ($\lambda$6580) absorption feature evolves from 12,720\,$\pm$\,220 km s$^{-1}$ to 10,670\,$\pm$\,180 km s$^{-1}$ (Fig~\ref{Fig11}). The measured C\,{\sc ii} $\lambda 6580$ line velocity is significantly lower than Si\,{\sc ii} $\lambda 6355$, with the $v_{\rm{C\,II}}/v_{\rm{Si\,II}}$ ratio varying from $\sim 0.78$ on day $-13.8$ to $\sim 0.91$ on day $-5.9$. This variation in the velocity ratio is dissimilar with the results of \cite{2011ApJ...732...30P}, who have shown the ratio to be constant over time, in a sample of 19 SNe Ia. Their sample also indicates the ratio to be $\sim 1$ within 10\%. The discrepancy between the photospheric velocity (as measured by Si\,{\sc ii} velocity) and the carbon velocity is explained by \cite{2011ApJ...732...30P} as being due to a clumpy carbon layer that is offset by an angle $\theta$ from the line of sight. The angle can be estimated if $v_{\rm{C\,II}} < v_{\rm{Si\,II}}$. The observed velocity ratios indicate $\theta$ to be $\sim 40^{\circ}$ on day $-13.8$ and $\sim 25^{\circ}$ on day $-5.9$. It is suggested that the change in the ratio (and angle from the line of sight) is indicative of an initial asymmetry that became more symmetric as the SN evolved to maximum, or a clumpiness that became more homogeneous as the SN ejecta evolved. Another possible explanation for the lower velocity ratio is mixing within the ejecta. The Ca\,{\sc ii} (H \& K) $\lambda$3951 absorption feature is seen to be around $\sim$\,22,000 km s$^{-1}$ at $-13.8$ d and always remains higher than Si\,{\sc ii} $\lambda$6355. 

\cite{2005ApJ...623.1011B} made a hierarchical cluster analysis of SN Ia based on both photometric ($\Delta m_{15}(B)$, $M_{B}$) and spectroscopic ($\dot{v}$, $v_{10}$(Si\,{\sc ii}), R(Si\,{\sc ii})$_{\rm{max}}$) parameters of SNe. Here, $\dot{v}$ is the average daily decline rate of expansion velocity of Si\,{\sc ii} $\lambda$6355 line, derived from the least square fit to the measurements taken between maximum and either the time the Si\,{\sc ii} feature disappears or the last available spectrum, whichever is earlier. The velocity of Si\,{\sc ii}, 10 days past maximum is denoted as $v_{10}$.

From the velocity evolution of SN~2017hpa, values of $\dot{v}$, $v_{10}$ and velocity at maximum are estimated as $127.9 \pm 6.1$ km s$^{-1}$d$^{-1}$, $\sim 8320$ km s$^{-1}$ and $\sim 9640$ km s$^{-1}$, respectively. With the measured value of $\dot{v}$, SN~2017hpa can be placed under the HVG group of \citet{2005ApJ...623.1011B} scheme. It is worth mentioning that the average value of $v_{10}$ for the HVG group is $12200 \pm 1100$ km s$^{-1}$ which is significantly higher than $v_{10}$ estimated for SN~2017hpa ($\sim 8320$ km s$^{-1}$). Based on the photometric and spectroscopic properties, SN~2017hpa can be placed under normal Ia with HVG.

The Si\,{\sc ii} $\lambda$6355 line velocity evolution is plotted along with other high-velocity gradient and normal SNe (see Fig.~\ref{Fig11}, right panel). For SN~2003du, and SN~2005cf, it is seen that the post maximum velocity evolution is flat. From the velocity evolution, sometimes the channel of explosion can be inferred.

\cite{2011ApJ...742...89F} used a linear fit to the data between $-6 \le t \le 10$~d. The slope gives the velocity gradient, and the offset ($v^{0}$) is the velocity at $B$-band maximum. Using this relation, the velocity gradient has been found to be $306.9 \pm 25.1$ km s$^{-1}$d$^{-1}$ and  $v^{0} \sim 9870$ km s$^{-1}$.  
Using the expansion velocity of Si\,{\sc ii} $\lambda$6355 at maximum \cite{2009ApJ...699L.139W} classified SNe Ia into two broad groups, High Velocity (HV, $v \ge 11,800$ km s$^{-1}$ and Normal Velocity ($10,600 \pm 400$ km s$^{-1}$). According to this, SN~2017hpa clearly falls in the Normal velocity group. 

\subsection{Spectral Fitting in the pre-maximum phase}

The spectral features in the pre-maximum spectra at phases $-$13.8 day and $-8.9$ day are identified using synthetic spectra generated using the {\small \it{syn++}}\footnote{\url{https://c3.lbl.gov/es/}} code (\citealt{2010ascl.soft10055P, 2011PASP..123..237T}). The synthetic spectra are also used to obtain an estimate of the velocity distribution. The code is based upon simple assumptions of spherical symmetry, homologous expansion, a sharp photosphere that emits a continuous blackbody spectrum, line formation by resonant scattering assuming Sobolev approximation. The synthetic spectrum consists of blended P-Cygni profiles superimposed on a continuum. The code has been extensively used for line identifications, estimating the velocity of the pseudophotosphere and constraining the velocity interval over which lines due to various ions are formed. For each ion introduced into the fitting, the optical depth at the photospheric velocity is a fitting parameter. The optical depth of other lines are calculated assuming Boltzmann excitation temperature ($T_{\rm{exc}}$). The optical depths of lines are taken to be exponentially decreasing function of velocity with the e-folding velocity (aux). At each epoch, the important fitting parameters are the velocity at the photosphere ($v_{\rm{phot}}$), the optical depth of the ion reference line ($\rm \tau$), the minimum and maximum velocity ($v_{\rm{min}}$ \& $v_{\rm{max}}$) imposed on the ions. When the minimum velocity exceeds the photosphere's velocity, then the ion is said to be detached. The blackbody fit temperature ($T_{\rm{BB}}$) represents the overall shape of the spectrum. 

The spectrum at $-13.8$ day is fit with photospheric velocity of $\sim 15,000$ km s$^{-1}$ and blackbody temperature of 10,000 K to match the slope of the observed spectrum. The temperature is consistent with that obtained by fitting a blackbody to the photometric spectral energy distribution (SED). The spectrum is fit with ions of IME's like C\,{\sc ii}, Mg\,{\sc ii}, Si\,{\sc ii} (PV \& HV), Ca\,{\sc ii} (PV \& HV) and IGE's like Fe\,{\sc ii} and Fe\,{\sc iii}. An excitation temperature of 10,000 K and a maximum velocity of 30,000 km s$^{-1}$ have been used for all the ions. To fit the Si\,{\sc ii} $\lambda$6355 line profile in the observed spectrum, a high velocity component at 22,000 km s$^{-1}$ has been introduced. From the fit to the spectrum, it is evident that both the C\,{\sc ii} $\lambda$6580 and Si\,{\sc ii} $\lambda$6355 features can be reproduced with a photospheric velocity of 15,000 km s$^{-1}$. This indicates that the line forming region of Si\,{\sc ii} and C\,{\sc ii} are moving at the same velocity. The Ca\,{\sc ii} (H \& K) profile is also fitted with two components, one at 17,000 km s$^{-1}$ and another, high-velocity component, with a velocity of 25,000 km s$^{-1}$.

\begin{figure}
\centering
\resizebox{\hsize}{!}{\includegraphics[width=\columnwidth]{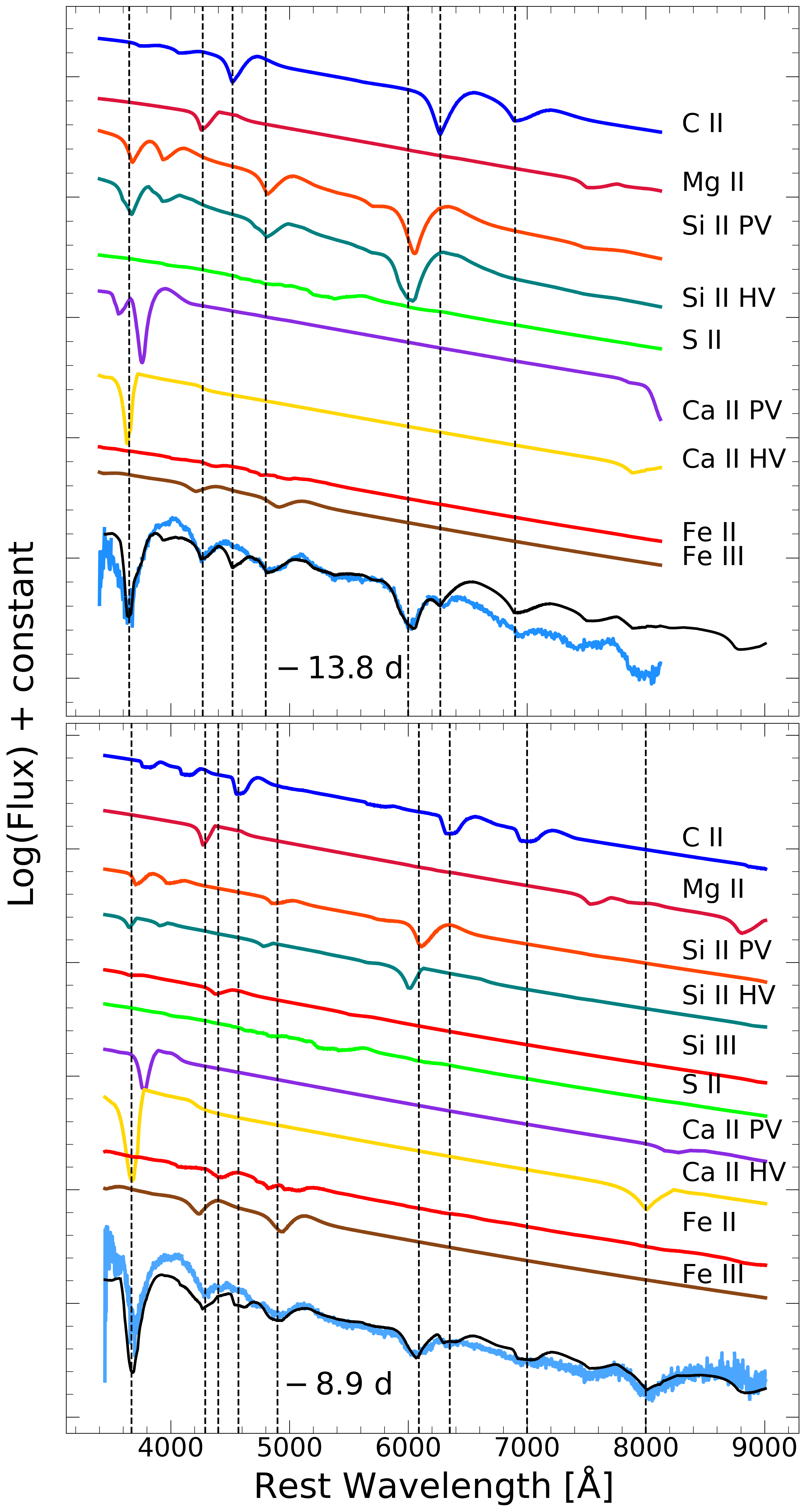}}
\caption{Dereddened and redshift corrected spectra of SN~2017hpa at $-13.8$~d and $-8.9$~d, plotted along with the synthetic spectra (black solid line) generated using {\small \it syn++}. The contributions from each element are shown.}
\label{Fig12}
\end{figure}

The spectrum at $-8.9$ d is fit with a photospheric velocity of 13,800 km s$^{-1}$ and a blackbody temperature of 11,000 K. In this phase also, the observed Si\,{\sc ii} $\lambda$6355 feature can be reproduced by a combination of the photospheric component and a high velocity component at 18,000 km s$^{-1}$. The C\,{\sc ii} feature is fit with a velocity of 13,800 km s$^{-1}$, similar to the photospheric component of Si\,{\sc ii}. Si\,{\sc iii} ($\lambda$4553, 4568, 4575) features are well reproduced with the photospheric velocity of 13,800 km s$^{-1}$. The spectral fits are shown in Fig.~\ref{Fig12}.
    
Most of the prominent features observed in the spectra of 2017hpa in the pre-maximum phase are reproduced in the {\small \it {syn++}} synthetic spectra. We note that certain components will require careful fitting in an iterative way. The details of the fit are provided in Table~\ref{tab:syn++}.
    
\begin{figure}
\centering
\resizebox{\hsize}{!}{\includegraphics[width=\columnwidth]{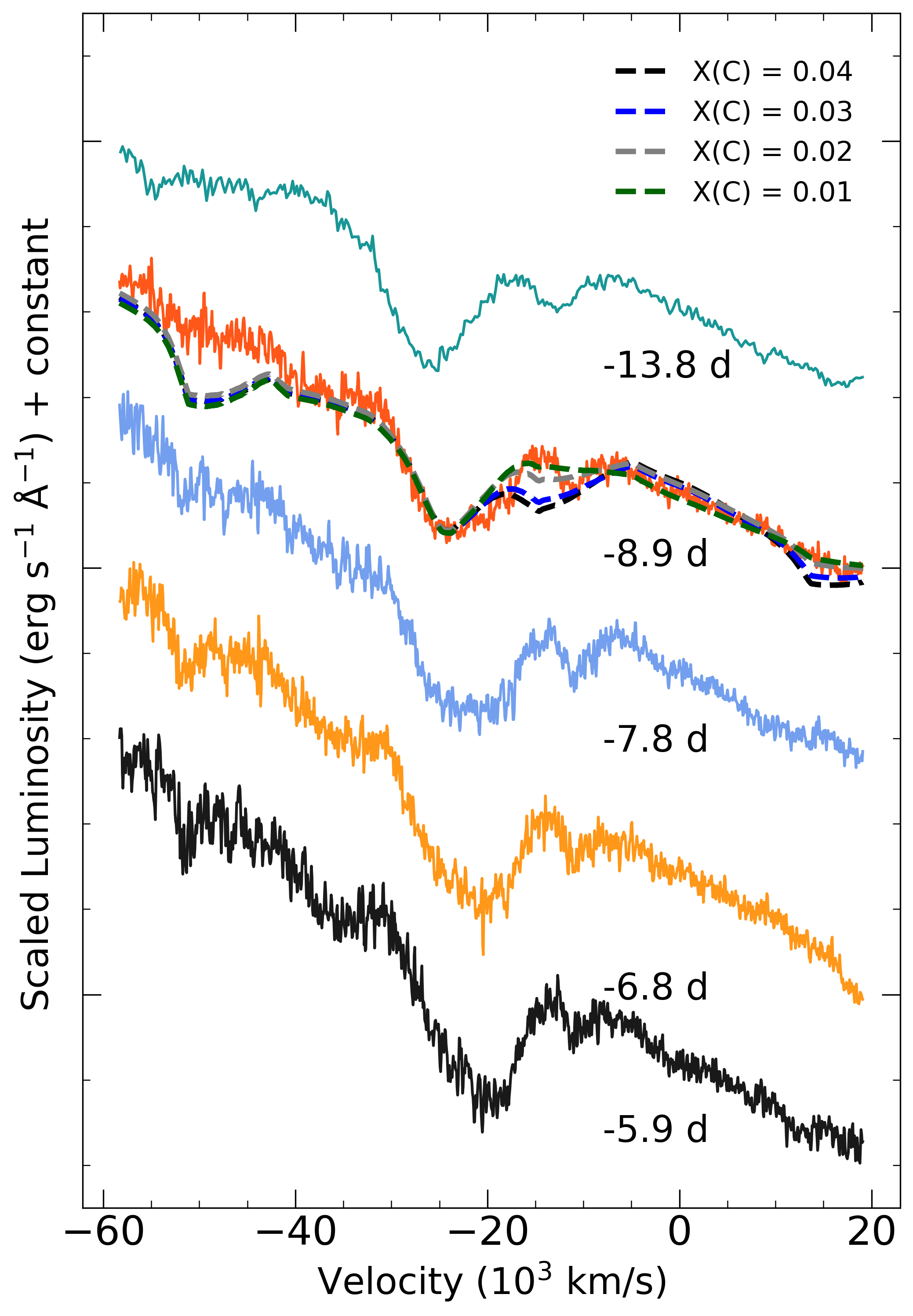}}
\caption{Velocity evolution of C\,{\sc ii} $\lambda$6580. The velocity is defined with respect to rest wavelength of  C\,{\sc ii} $\lambda$6580. The dotted solid lines indicate {\small \it{TARDIS}} fit to the observed spectra with different mass fraction for Carbon.}
\label{Fig13}
\end{figure}

To study the explosion mechanism and to put a constraint on the mass of unburned C present in the ejecta, we applied the one-dimensional spectrum synthesis code {\small \it{TARDIS}} \citep{2014MNRAS.440..387K} to our spectrum at $-8.9$~d. The fit parameters of {\small \it{TARDIS}} are listed in Table~\ref{tab:Tardis}. {\small \it{TARDIS}} takes as input the luminosity of the SN, the time since the explosion, a density, velocity and abundance profile. {\small \it{TARDIS}} assumes a single, sharp photosphere is emitting a quasi-blackbody continuum between the optically thick and thin regions of the ejecta. For a uniform abundance, the region above the photosphere is divided into multiple, spherically symmetric cells. By specifying the mass fraction of the elements as inputs in these cells, {\small \it{TARDIS}} generates a synthetic spectrum by calculating the ionisation and excitation states. In this work, we model the spectrum by using a uniform abundance of elements, and a W7-like (\citealt{1984ApJ...286..644N, 1985ApJ...294..619B}) density profile. The observed spectrum does not have prominent lines due to IGE's; hence we consider only C, O, Mg, Si and Ca for fitting the observed spectrum. 
By using a uniform composition in all the cells, a satisfactory fit to the observed spectrum was not possible in the entire wavelength range. Since we observe strong C feature in the pre-maximum spectra around $\lambda$6580, we tried to put a constraint on the mass of unburned C. An attempt is made to fit the region around Si\,{\sc ii} and C\,{\sc ii} by varying the mass fraction of C from 0.01 to 0.04  and keeping the mass fraction of other elements the same (see Fig.~\ref{Fig13}). It is found that for the mass fraction of C more than 0.03, the C\,{\sc ii} line blends with Si\,{\sc ii}, hence we put an upper limit to the C mass fraction as 0.03. By integrating the contribution from all cells, we find $\sim 0.019 M_\odot$ of C in the outer regions of the ejecta, which is consistent with the mass estimate by \cite{2011MNRAS.410.1725T} for SN~2003du. The estimated carbon mass is significantly lower than the mass expected in a deflagration model ($0.049\,M_\odot$). This could be due to mixing in the ejecta.

\section{Summary}
\label{Summary}

The properties of SN~2017hpa make it an interesting object. From the photometric and spectroscopic properties, it can be classified as a normal type Ia object with $\Delta$m$_{15}$(B)\,=\,$0.98\pm0.16$ and M$^{max}_{B}$\,=\,--19.45\,$\pm$\,0.15 with a distinct secondary maximum in $I$ band. With standard light curve fitting methods like MLCS2k2 and SALT2, the distance modulus is estimated to be $34.08\pm0.09$ with a host reddening of $E(B-V)=0.08$ and an R\(_{V}\) of 1.9. The $(B-V)$ colour evolution is bluer than the comparison SNe, and from the $(uvw1-uvv$) colour evolution, SN~2017hpa can be placed under NUV-blue group. By fitting the bolometric light curve with the radiation diffusion model, we derive nickel mass ($M_{\rm{Ni}}$) of $0.61 M_\odot$ and ejecta mass of $1.10 M_\odot$.

The pre-maximum spectral sequence of SN~2017hpa shows a relatively featureless continuum. The spectral evolution shows strong C\,{\sc ii} $\lambda$6580 line, which can be seen until 5.9 days before maximum. The observed velocity of C\,{\sc ii} is lower than Si\,{\sc ii} $\lambda$6355. However the \textit{syn++} fit to the observed spectra shows that both the C\,{\sc ii} and Si\,{\sc ii} line forming regions are moving with the photospheric velocity. The Si\,{\sc ii} $\lambda$6355 shows rapid evolution in velocity and the velocity gradient measured is $127.9 \pm 6.1$ km s$^{-1}$ d$^{-1}$ which places SN~2017hpa among the HVG objects according to the Benetti classification scheme. The velocity measured at the maximum is $\sim 9643$ km s$^{-1}$ which is clearly within the range for normal velocity SNe Ia. Typically, C\,{\sc ii} is associated with the low-velocity gradient group \citep{2012AJ....143..126B}. The $(B-V)$ colour at maximum is redder for the HVG subtype \citep{2018ApJ...864L..35S}. The observed properties of SN~2017hpa are strikingly opposite to this. {\small \it{TARDIS}} model fit to our spectrum requires unburned C mass to be $\sim 0.019\, M_\odot$, lower than that expected by deflagration alone. Detailed three-dimensional modelling may confirm the exact nature of SN~2017hpa. 

\section*{Acknowledgements}

We thank the referee for going through the manuscript carefully and providing constructive suggestions which helped in improving the content of the paper. We thank the staff of IAO, Hanle and CREST, Hosakote that made the observations possible. The facilities at IAO and CREST are operated by the Indian Institute of Astrophysics, Bangalore. We also thank the observers who shared their valuable time in Target of Opportunity (ToO) observations during the initial follow up. This work has made use of the NASA Astrophysics Data System\footnote{\url{https://ui.adsabs.harvard.edu/}} (ADS), the NASA/IPAC extragalactic database\footnote{\url{https://ned.ipac.caltech.edu/}} (NED) and NASA/IPAC Infrared Science Archive (IRSA)\footnote{\url{https://irsa.ipac.caltech.edu/applications/DUST/}} which is operated by the Jet Propulsion Laboratory, California Institute of Technology. We acknowledge, Weizmann Interactive Supernova Data REPository\footnote{\url{https://wiserep.weizmann.ac.il/}} (WISeREP, \cite{2012PASP..124..668Y}. The research has made use of the data obtained from the High Energy Astrophysics Science Archive Research Center (HEASARC) \footnote{\url{https://heasarc.gsfc.nasa.gov/}}, a facility of the Astrophysics Science Division at NASA/GSFC and of the Smithsonian Astrophysical Observatory's High Energy Astrophysics Division. This research made use of \textit{Tardis}, a community-developed software package for spectral synthesis in supernovae (\citealt{2014MNRAS.440..387K, 2019zndo...2590539K}). The development of \textit{Tardis} received support from the Google Summer of Code initiative and from ESA's Summer of Code in Space program. \textit{Tardis} makes extensive use of \textit{Astropy} and \textit{PyNE}. The reduction of photometric and spectroscopic data has been greatly facilitated by \textsc{RedPipe}\footnote{\url{https://github.com/sPaMFouR/RedPipe}}.The analysis has made use of the following software and packages - (i) {\small \it{Image Reduction and Analysis Facility (IRAF)}}, \cite{1993ASPC...52..173T}. (ii) {\small \it{PyRAF}}, \cite{2012ascl.soft07011S}. (iii) {\small \it{NumPy}}, \cite{van2011numpy}, (iv) {\small \it{Matplotlib}}, \cite{Hunter:2007}, (v) {\small \it{SciPy}}, \cite{2020SciPy-NMeth}, (vi) {\small \it{pandas}}, \cite{reback2020pandas}, (vii) {\small \it{astropy}}, \cite{2013A&A...558A..33A}, (viii) {\small \it{emcee}}, \cite{2013ascl.soft03002F}, (ix) {\small \it{syn++}}, \cite{2011PASP..123..237T}, (x) {\small \it{TARDIS}}, \cite{2014MNRAS.440..387K}. 

%%%%%%%%%%%%%%%%%%%%%%%%%%%%%%%%%%%%%%%%%%%%%%%%%%
\section*{Data Availability}

The UV and Optical magnitudes are tabulated in Table~\ref{tab:uvlog} and \ref{tab:photlog},  respectively. The log of spectroscopic data is provided in Table~\ref{tab:speclog}. The $Swift-UVOT$ data are available at \url{https://www.swift.ac.uk/swift_portal/}. The spectroscopic data used in this work will be made available on request.
%%%%%%%%%%%%%%%%%%%% REFERENCES %%%%%%%%%%%%%%%%%%

% The best way to enter references is to use BibTeX:

\bibliographystyle{mnras}
\bibliography{biblio} % if your bibtex file is called example.bib

\appendix

\section{Tables}
\begin{table*}
\centering
\renewcommand{\arraystretch}{1.1}
\caption{$UBVRI$ magnitudes of local standards in the field of SN 2017hpa.}
\label{tab:standardlog}
\begin{tabular}{c c c c c c}
\hline \hline
      ID         &         $U$               &    $B$                &    $V$                &    $R$                &    $I$       \\
 \hline \noalign{\smallskip}
 1      &       18.45$\,\pm\,$0.03   &   16.99$\,\pm\,$0.01  &   15.66$\,\pm\,$0.02  &   14.85$\,\pm\,$0.01  &   14.16$\,\pm\,$0.03  \\
 2      &       17.49$\,\pm\,$0.03   &   17.13$\,\pm\,$0.02  &   16.26$\,\pm\,$0.02  &   15.75$\,\pm\,$0.01  &   15.27$\,\pm\,$0.02  \\
 3      &       18.13$\,\pm\,$0.04   &   18.24$\,\pm\,$0.02  &   17.63$\,\pm\,$0.01  &   17.20$\,\pm\,$0.02  &   16.79$\,\pm\,$0.02  \\
 4      &       18.02$\,\pm\,$0.03   &   17.52$\,\pm\,$0.02  &   16.61$\,\pm\,$0.01  &   16.07$\,\pm\,$0.02  &   15.54$\,\pm\,$0.02  \\
 5      &       16.89$\,\pm\,$0.03   &   16.81$\,\pm\,$0.02 &    16.04$\,\pm\,$0.01  &   15.56$\,\pm\,$0.01  &   15.06$\,\pm\,$0.02  \\
 6      &       18.79$\,\pm\,$0.02   &   18.25$\,\pm\,$0.03  &   17.38$\,\pm\,$0.01  &   16.86$\,\pm\,$0.01  &   16.38$\,\pm\,$0.02  \\
 7      &       18.62$\,\pm\,$0.04   &   17.76$\,\pm\,$0.02  &   16.72$\,\pm\,$0.01  &   16.08$\,\pm\,$0.02  &   15.52$\,\pm\,$0.02  \\
 
\noalign{\smallskip} \hline
%\multicolumn{3}{l}{$^*$\footnotesize{Time since $B$-band maximum (JD 2458066.3)}.}
\end{tabular}
\end{table*}
\begin{table*}
\centering
\renewcommand{\arraystretch}{1.1}
\caption{UV-Optical photometry of SN 2017hpa with {\it Swift-UVOT}.}
\label{tab:uvlog}
\begin{tabular}{lcccccccc}
\hline \hline
 Date & JD & Phase$^*$ & $uvw1$ & $u$ & $b$ & $v$\\
  (yyyy-mm-dd)   &  (2458000+)  & (d)  & (mag)  & (mag)  & (mag)  & (mag)\\
\hline \noalign{\smallskip}
2017-10-26 & 53.25 & $-$13.05 & 18.37$\pm$0.16 &  17.59$\pm$0.10  & 17.09$\pm$0.05 &  16.63$\pm$0.06 \\
2017-10-28 & 54.58 & $-$11.72 & 18.35$\pm$0.24 &  16.91$\pm$0.12  & 16.74$\pm$0.07 &  16.49$\pm$0.12 \\
2017-10-30 & 56.98 & $-$9.32  & 18.09$\pm$0.23 &  16.35$\pm$0.09  & 16.38$\pm$0.07 &  16.17$\pm$0.11 \\
2017-11-03 & 60.89 & $-$5.41  & 17.00$\pm$0.10 &  15.38$\pm$0.04  & 15.62$\pm$0.04 &  15.66$\pm$0.06 \\
2017-11-09 & 67.48 &    1.18  & 16.80$\pm$0.13 &  15.37$\pm$0.07  & 15.46$\pm$0.05 &  15.31$\pm$0.08 \\
2017-11-11 & 69.26 &    2.97  & 17.11$\pm$0.15 &  15.46$\pm$0.05  &     ---        &  15.38$\pm$0.09 \\
2017-11-13 & 71.11 &    4.82  & 17.19$\pm$0.09 &  15.61$\pm$0.09  & 15.54$\pm$0.04 &  15.31$\pm$0.05 \\
2017-11-15 & 73.32 &    7.02  & 17.33$\pm$0.19 &  15.82$\pm$0.08  & 15.70$\pm$0.06 &  15.36$\pm$0.09 \\
2017-11-17 & 75.18 &    8.88  & 17.86$\pm$0.25 &  15.95$\pm$0.08  & 15.80$\pm$0.06 &  15.39$\pm$0.05 \\
2017-11-23 & 81.08 &   14.78  & 17.97$\pm$0.16 &  16.77$\pm$0.09  & 16.39$\pm$0.05 &  15.81$\pm$0.08 \\
2017-11-24 & 82.35 &   16.05  & 18.31$\pm$0.19 &  16.93$\pm$0.09  & 16.53$\pm$0.06 &  15.84$\pm$0.07 \\
2017-11-27 & 85.27 &   18.97  & 18.59$\pm$0.21 &  17.32$\pm$0.12  & 16.91$\pm$0.07 &  16.01$\pm$0.07 \\
2017-12-01 & 88.86 &   22.56  & 19.36$\pm$0.56 &  17.46$\pm$0.16  & 17.21$\pm$0.10 &  16.20$\pm$0.10 \\
2017-12-07 & 95.43 &   29.13  &     ---        &  18.99$\pm$0.45  & 17.65$\pm$0.11 &  16.48$\pm$0.10 \\
\noalign{\smallskip} \hline
\multicolumn{3}{l}{$^*$\footnotesize{Time since $B$-band maximum (JD 2458066.3)}.}
\end{tabular}			    
\end{table*}
\begin{table*}
\centering
\renewcommand{\arraystretch}{1.1}
\caption{Optical photometry of SN2017hpa from HCT.}
\label{tab:photlog}
\begin{tabular}{c c c c c c c c}
\hline \hline
       Date     &      JD       & Phase$^*$ &     $U$               &    $B$                &    $V$                &    $R$                &    $I$                \\
(yyyy-mm-dd)    & (2458000+)    &   (d)     &   (mag)               &   (mag)               &   (mag)               &   (mag)               &   (mag)               \\
 \hline \noalign{\smallskip}
 2017-10-31     &    58.45      &   --7.8   &       15.68$\,\pm\,$0.13   &   15.98$\,\pm\,$0.03  &   15.95$\,\pm\,$0.02  &   15.82$\,\pm\,$0.02  &   15.75$\,\pm\,$0.03  \\
 2017-11-01     &    59.43      &   --6.8   &       15.49$\,\pm\,$0.15   &   15.77$\,\pm\,$0.03  &   15.82$\,\pm\,$0.02  &   15.69$\,\pm\,$0.02  &   15.63$\,\pm\,$0.02  \\
 2017-11-02     &    60.37      &   --5.9   &       15.37$\,\pm\,$0.15   &   15.63$\,\pm\,$0.06  &   15.75$\,\pm\,$0.03  &   15.54$\,\pm\,$0.03  &   15.52$\,\pm\,$0.02  \\
 2017-11-09     &    67.19      &    0.9    &       15.26$\,\pm\,$0.14   &   15.41$\,\pm\,$0.02  &   15.39$\,\pm\,$0.01  &   15.31$\,\pm\,$0.02  &   15.43$\,\pm\,$0.02  \\
 2017-11-10     &    68.21      &    2.0    &       15.36$\,\pm\,$0.19   &   15.44$\,\pm\,$0.03  &   15.40$\,\pm\,$0.01  &   15.30$\,\pm\,$0.01  &   15.45$\,\pm\,$0.01  \\
 2017-11-15     &    73.27      &    7.0    &       15.69$\,\pm\,$0.15   &   15.68$\,\pm\,$0.03  &   15.39$\,\pm\,$0.01  &   15.39$\,\pm\,$0.01  &   15.61$\,\pm\,$0.01  \\
 2017-11-19     &    77.43      &   11.1    &               ---          &   16.04$\,\pm\,$0.04  &   15.48$\,\pm\,$0.01  &   15.63$\,\pm\,$0.02  &   16.02$\,\pm\,$0.04  \\
 2017-11-22     &    80.28      &   13.9    &       16.57$\,\pm\,$0.09   &   16.36$\,\pm\,$0.02  &   15.75$\,\pm\,$0.01  &   15.79$\,\pm\,$0.01  &   15.96$\,\pm\,$0.02  \\
 2017-11-29     &    87.36      &   21.1    &               ---          &   17.23$\,\pm\,$0.02  &   16.26$\,\pm\,$0.02  &   15.98$\,\pm\,$0.01  &   15.83$\,\pm\,$0.02  \\
 2017-12-17     &    105.24     &   38.9    &               ---          &   18.38$\,\pm\,$0.02  &   17.18$\,\pm\,$0.02  &   16.74$\,\pm\,$0.01  &   16.30$\,\pm\,$0.02  \\
 2017-12-26     &    114.21     &   47.9    &               ---          &   18.60$\,\pm\,$0.03  &   17.48$\,\pm\,$0.02  &   17.12$\,\pm\,$0.01  &   16.76$\,\pm\,$0.02  \\ 
 2017-12-29     &    117.24     &   50.9    &               ---          &   18.62$\,\pm\,$0.04  &   17.54$\,\pm\,$0.02  &   17.20$\,\pm\,$0.02  &   16.92$\,\pm\,$0.02  \\
 2018-01-05     &    124.09     &   57.8    &               ---          &          ---          &   17.78$\,\pm\,$0.02  &   17.48$\,\pm\,$0.01  &   17.33$\,\pm\,$0.03  \\
 2018-01-21     &    140.12     &   73.8    &       19.55$\,\pm\,$0.25   &   19.02$\,\pm\,$0.03  &   18.12$\,\pm\,$0.03  &   17.94$\,\pm\,$0.02  &   17.97$\,\pm\,$0.03  \\
 2018-02-02     &    152.11     &   85.8    &               ---          &          ---          &   18.37$\,\pm\,$0.02  &   18.25$\,\pm\,$0.02  &          ---  \\
 2018-02-03     &    153.14     &   86.9    &       19.95$\,\pm\,$0.19   &   19.08$\,\pm\,$0.07  &   18.40$\,\pm\,$0.02  &   18.24$\,\pm\,$0.03  &   18.26$\,\pm\,$0.04  \\
 2018-02-09     &    159.27     &   93.0    &               ---          &          ---          &   18.58$\,\pm\,$0.03  &   18.48$\,\pm\,$0.02  &   18.63$\,\pm\,$0.04  \\ 
\noalign{\smallskip} \hline
\multicolumn{3}{l}{$^*$\footnotesize{Time since $B$-band maximum (JD 2458066.3)}.}
\end{tabular}
\end{table*}
\begin{table*}
\centering
\setlength{\tabcolsep}{3pt}
\caption{Log of spectroscopic observations of SN2017hpa from HCT.}
\label{tab:speclog}
\begin{tabular}{c c c l}
\hline \hline
     Date     &     JD      & Phase$^*$ &        Range           \\
 (yyyy-mm-dd) & (2458000+)  &  (d)      &        (\AA)           \\
 \hline \noalign{\smallskip}
 2017-10-30   &   57.43     &  --8.9     & 3500-7800; 5200-9100   \\
 2017-10-31   &   58.46     &  --7.8     & 3500-7800; 5200-9100   \\
 2017-11-01   &   59.47     &  --6.7     & 3500-7800            \\
 2017-11-02   &   60.39     &  --5.9     & 3500-7800; 5200-9100   \\
 2017-11-09   &   67.22     &   0.9     & 3500-7800; 5200-9100   \\
 2017-11-10   &   68.23     &   1.9     & 3500-7800; 5200-9100   \\
 2017-11-19   &   77.36     &   11.1    & 3500-7800; 5200-9100   \\
 2017-11-22   &   80.30     &   14.1    & 3500-7800; 5200-9100   \\
 2017-11-29   &   87.23     &   20.9    & 3500-7800; 5200-9100   \\
 2017-12-01   &   89.31     &   23.1    & 3500-7800; 5200-9100   \\
 2017-12-05   &   93.19     &   26.9    & 3500-7800; 5200-9100   \\
 2017-12-17   &  105.26     &   39.0    & 3500-7800; 5200-9100   \\
 2017-12-26   &  114.24     &   47.9    & 3500-7800; 5200-9100   \\
 2017-12-29   &  117.25     &   51.0    & 3500-7800; 5200-9100   \\
 2018-01-05   &  124.13     &   57.9    & 3500-7800; 5200-9100   \\
 2018-01-21   &  140.22     &   73.9    & 3500-7800; 5200-9100   \\
 2018-01-30   &  149.17     &   82.9    & 3500-7800; 5100-9100   \\
 2018-02-03   &  153.07     &   86.8    & 3500-7800             \\
 2018-02-09   &  159.14     &   92.8    & 3500-7800; 5200-9100   \\
 2018-02-13   &  163.13     &   96.8    & 3500-7800             \\
 2018-02-20   &  170.07     &  103.8    & 3500-7800             \\
 2018-02-25   &  175.06     &  108.8    & 3500-7800             \\
\noalign{\smallskip} \hline
\multicolumn{3}{l}{$^*$\footnotesize{Time since $B$-band maximum (JD 2458066.3)}.}
\end{tabular}
\end{table*}
\begin{table*}
\centering
\caption{{\small \textit{syn++}} fit to the pre-maximum spectra of SN~2017hpa.}
\label{tab:syn++}
\begin{tabular}{c c c c}
\hline
\hline
\noalign{\smallskip}
Phase$^*$: $-$ 13.8 d  & v\(_{phot}\): 15,000 km s$^{-1}$  & v\(_{max}\): 30,000 km s$^{-1}$ &  T\(_{BB}\): 10,000 K \\
\hline
\end{tabular}
\begin{tabular}{c c c c c c c c c c c}
Parameters & C\,{\sc ii} & Mg\,{\sc ii} & Si\,{\sc ii}$\rm _{PV}$ & Si\,{\sc ii}$\rm _{HV}$ & S\,{\sc ii} & Ca\,{\sc ii}$\rm _{PV}$ & Ca\,{\sc ii}$\rm _{HV}$ & Fe\,{\sc ii} & Fe\,{\sc iii} \\
log (tau) &  -0.5 & 0.5 & 0.9 & 1.5 & 0.2 & 1.0 & 2.3 & 1.0 & -0.8 \\
v$_{min}$ (\(\times\) 10$\rm ^{3}$) km s$^{-1}$ & 15.0 & 16.0 & 15.0 & 22.0 & 15.0 & 17.0 & 25.0 & 17.0 & 15.0 \\
aux (\(\times\) 10$\rm ^{3}$) km s$^{-1}$ & 2.1 & 1.8 & 2.6 & 2.45 & 1.0 & 2.0 & 2.7 & 0.8 & 3.2 \\
T$_{exc}$ (\(\times\) 10$\rm ^{3}$) K & 10 & 10 & 10 & 10 & 10 & 10 & 10 & 10 & 10 \\
\hline
\hline
\end{tabular}
\begin{tabular}{c c c c}
\hline
\hline
\noalign{\smallskip}
Phase$^*$: $-$ 8.9 d  & v\(_{phot}\): 13,800 km s$^{-1}$  & v\(_{max}\): 30,000 km s$^{-1}$ &  T\(_{BB}\): 11,000 K \\
\hline
\end{tabular}
\begin{tabular}{c c c c c c c c c c c c}
Parameters & C\,{\sc ii} & Mg\,{\sc ii} & Si\,{\sc ii}$\rm _{PV}$ & Si\,{\sc ii}$\rm _{HV}$ & Si\,{\sc iii} & S\,{\sc ii} & Ca\,{\sc ii}$\rm _{PV}$ & Ca\,{\sc ii}$\rm _{HV}$ & Fe\,{\sc ii} & Fe\,{\sc iii} \\
log (tau) &  0.2 & 0.2 & 0.03 & 0.5 & -0.6 & -0.8 & 0.5 & 1.5 & -0.5 & -0.5 \\
v$_{min}$ (\(\times\) 10$\rm ^{3}$) km s$^{-1}$ & 13.8 & 15.0 & 13.8 & 18.0 & 13.8 & 13.8 & 14.0 & 20.0 & 13.8 & 13.8\\
aux (\(\times\) 10$\rm ^{3}$) km s$^{-1}$ & 0.4 & 1.7 & 3.5 & 2.5 & 2.5 & 0.8 & 2.0 & 4.0 & 1.0 & 3.0 \\
T$_{exc}$ (\(\times\) 10$\rm ^{3}$) K & 10 & 10 & 10 & 10 & 10 & 10 & 10 & 10 & 10 & 10 \\
\noalign{\smallskip} 
\hline
\multicolumn{1}{l}{$^*$\footnotesize{Time since B-band maximum (JD 2458066.3)}.}
\end{tabular}

\end{table*}
\begin{table*}
\centering
\renewcommand{\arraystretch}{1.1}
\caption{{\small \textit{TARDIS}} fit to the pre-maximum spectra of SN~2017hpa.}
\label{tab:Tardis}
\begin{tabular}{c c c c c}
\hline
\hline
\noalign{\smallskip}
Phase$^*$: $-$ 8.9 d & t\(_{exp}\)$^*$: 8.0 day & v\(_{inner}\)$^*$: 13,000 km s$^{-1}$  & v\(_{outer}\): 23,000 km s$^{-1}$ &  Luminosity (log L\(_{\odot}\)): 9.37 \\
\hline
\end{tabular}
\begin{tabular}{c c c c c c }
Elements    &     C  &    O    &    Mg   &   Si   &   Ca    \\
Abundances  &  0.03  &   0.26  &   0.10  &  0.53  &   0.08  \\
\noalign{\smallskip} \hline
\multicolumn{1}{l}{$^*$\footnotesize{Time since B-band maximum (JD 2458066.3)}.} \\
\multicolumn{2}{l}{$^*$\footnotesize{Time since explosion (JD 2458049.41)}.} \\ 
\multicolumn{3}{l}{$^*$\footnotesize{velocity at the inner boundary of the photosphere.}} 
\end{tabular}
\end{table*}
% Alternatively you could enter them by hand, like this:
% This method is tedious and prone to error if you have lots of references
%\begin{thebibliography}{99}
%\bibitem[\protect\citeauthoryear{Author}{2012}]{Author2012}
%Author A.~N., 2013, Journal of Improbable Astronomy, 1, 1
%\bibitem[\protect\citeauthoryear{Others}{2013}]{Others2013}
%Others S., 2012, Journal of Interesting Stuff, 17, 198
%\end{thebibliography}

%%%%%%%%%%%%%%%%%%%%%%%%%%%%%%%%%%%%%%%%%%%%%%%%%%

%%%%%%%%%%%%%%%%% APPENDICES %%%%%%%%%%%%%%%%%%%%%

%If you want to present additional material which would interrupt the flow of the main paper, it can be placed in an Appendix which appears after the list of references.

%%%%%%%%%%%%%%%%%%%%%%%%%%%%%%%%%%%%%%%%%%%%%%%%%%

% Don't change these lines
%\bsp	% typesetting comment
\label{lastpage}
\end{document}